\documentclass[10pt,twocolumn,letterpaper]{article}

\usepackage[pagenumbers]{cvpr}      
\usepackage[most]{tcolorbox}
\usepackage{xcolor,graphicx,multirow}
\definecolor{imgbg}{RGB}{248,249,251}
\definecolor{imgframe}{gray}{0.65}
\newlength{\colw}
\newcommand{\colgap}{0pt}                            
\newlength{\rowlabelw}

\newcommand{\colstack}[3]{%
  \begin{tcolorbox}[enhanced,
    equal height group=cols,    
    colback=imgbg, colframe=imgframe,
    boxrule=0.5pt, sharp corners, boxsep=0pt,
    left=2pt, right=2pt, top=2pt, bottom=2pt,
    width=\colw, valign=center]
    \includegraphics[width=\linewidth]{#1}\par\vspace{0pt}%
    \includegraphics[width=\linewidth]{#2}\par\vspace{0pt}%
    \includegraphics[width=\linewidth]{#3}%
  \end{tcolorbox}%
}

\definecolor{cvprblue}{rgb}{0.21,0.49,0.74}
\usepackage[pagebackref,breaklinks,colorlinks,allcolors=cvprblue]{hyperref}

\def\paperID{40420} 
\def\confName{CVPR}
\def\confYear{2026}

\title{A Geometric Algebra-Informed 3DGS Framework for Wireless Channel Prediction}

\author{
Jingzhou Shen \qquad Tianya Zhao \qquad Xuyu Wang\footnotemark\\
Knight Foundation School of Computing and Information Sciences\\
Florida International University\\
{\tt\small \{jshen020, tzhao010, xuywang\}@fiu.edu}
}

\begin{document}
\maketitle

\renewcommand{\thefootnote}{*}
\begin{NoHyper}
\footnotetext{The corresponding author is Xuyu Wang (xuywang@fiu.edu).}
\end{NoHyper}
\renewcommand{\thefootnote}{\arabic{footnote}}

\begin{abstract}
In this paper, we introduce Geometric Algebra–Informed 3D Gaussian Splatting (GAI-GS), a framework for wireless modeling that couples 3D Gaussian splatting with a geometric algebra–based attention mechanism to explicitly model ray–object interactions in complex propagation environments. GAI-GS encodes joint spatial–electromagnetic (EM) relations into token representations, enabling scene-level aggregation within a unified, end-to-end neural architecture. This design grounds wireless ray propagation in electromagnetic principles, allowing token interactions to model key effects such as multipath, attenuation, and reflection/diffraction. Through extensive evaluations on multiple real-world indoor datasets, GAI-GS consistently surpasses current baselines across various wireless tasks.
\end{abstract}    
\section{Introduction}
\label{sec:intro}

Modern society has witnessed an unprecedented integration of connected devices into every aspect of daily life. From smart sensors and wearable technology to autonomous systems and Internet of Things (IoT) devices, wireless communication forms an intricate web that underpins critical infrastructure and personal conveniences. This transformation has elevated wireless channel modeling, which characterizes electromagnetic wave propagation in diverse environments, to a fundamental challenge in telecommunications. Wireless channel modeling captures complex phenomena such as signal attenuation, reflection, diffraction, and scattering~\cite{7109864, wang2020indoor}, providing essential insights for effective network design, resource allocation, wireless localization, and quality of service optimization in increasingly dense and heterogeneous wireless networks~\cite{wang2018survey, wang2022adversarial}.

At the core of wireless communications lies the physics of electromagnetic wave propagation governed by Maxwell’s equations~\cite{7152831, rappaport2017overview}. Directly solving these equations in realistic environments is intractable due to incomplete boundary conditions and complex geometries~\cite{almers2007survey}, motivating approximate modeling strategies. Classical approaches fall into probabilistic, deterministic, and, more recently, neural modeling~\cite{huang2018big, he2018clustering, 8438326}. Probabilistic models use empirical statistics to relate received signal strength to distance and a few coarse parameters; they are efficient but provide limited spatial detail and cannot accurately resolve angle-of-arrival distributions. Deterministic models leverage physical optics and CAD-like environment descriptions to generate richer propagation characteristics~\cite{6387266}, yet still struggle to capture fine-grained material and structural complexity in real-world scenes~\cite{imoize2021standard}.

On the other hand, machine learning models~\cite{aldossari2019machine, yang2019generative, 7792374} bypass rigid statistical assumptions and simplified electromagnetic approximations, instead inferring relationships between scene geometry and signal behavior from data~\cite{8422221,9771907, 10.1145/3300061.3345438}. Neural radiance fields~(NeRF) extend this idea by learning continuous volumetric functions that map spatial coordinates to propagation-related quantities under measurement supervision~\cite{mildenhall2021nerf}. NeRF$^2$~\cite{10.1145/3570361.3592527} adapts this framework to wireless channels by jointly encoding geometry and signal characteristics, while NeWRF~\cite{10.5555/3692070.3693415} incorporates electromagnetic priors into volumetric rendering to enhance spatial consistency. Recent works further apply NeRF-style models to wireless field reconstruction and generalizable channel prediction~\cite{jia2025neuralreflectancefieldsradiofrequency, 10.1145/3678572, chen2025radio, 11206262}. Despite their accuracy, NeRF-based approaches remain computationally demanding for real-time or large-scale deployment. 3D Gaussian Splatting~(3D-GS)~\cite{kerbl20233d, zhang2024fregs, jiang2025geometry, chen2024mvsplat} represents a scene as an explicit set of anisotropic 3D Gaussians, enabling high quality and real-time view synthesis. Current 3D-GS models~\cite{11044513, wen2024neural} in the wireless domain tackle the challenge of accurately and efficiently reconstructing high-resolution spatial channel characteristics from sparse measurements in complex environments, enabling fast, site-specific wireless digital twins and downstream tasks.

However, existing wireless 3D-GS methods treat signal propagation as purely data-driven regression and overlook critical physical interactions between electromagnetic rays and environmental geometry. These approaches directly learn the mapping from spatial coordinates to signal strength without explicitly modeling ray-object interactions such as reflection, refraction, and diffraction at material boundaries. By neglecting the geometric properties of obstacles and their electromagnetic characteristics, these methods fail to capture the fundamental physics governing wave propagation. Therefore, we propose GAI-GS, a novel multi-view framework that effectively integrates geometric algebra~(GA)~\cite{dorst2022guided, Roelfs_2023} with Euclidean algebra. We introduce a specialized tokenizer with multiple algebraic embeddings to capture ray--object interactions from local to global scales. To better reflect how scene-level context is formed from the Gaussian representation, we explicitly describe the tokenizer instantiation using a subset of high-opacity Gaussian primitives as representative anchors. In addition, we clarify how the outputs of the scene mapping network are incorporated into Gaussian attributes through residual parameterization, enabling the learned representations to adapt to transmitter-dependent propagation conditions. Our contributions can be summarized as follows:

\begin{itemize}
    \item We propose the first geometric algebra–based 3D-GS framework for wireless channel modeling. By leveraging the unique mathematical structure of geometric algebra to capture local scattering patterns, our framework provides a comprehensive representation of electromagnetic wave propagation characteristics.
    \item We design a unified multi-view embedding architecture that combines Euclidean and geometric algebra representations. By implicitly learning ray-object interaction patterns along propagation trajectories, our approach encodes physically meaningful signal characteristics while leveraging the complementary strengths of geometric and Euclidean representations for improved performance.
    \item Our method achieves superior performance in multiple wireless datasets, outperforming existing baselines. Additionally, we release a custom-built dataset to support and advance future research in this domain. The dataset is available at: \url{https://huggingface.co/datasets/NorahCS/GAT-series_Dataset}.
\end{itemize}

\section{Preliminaries}
\label{sec:pre}

\subsection{Wireless Signal Representation}
\label{subsec:WCP}

In wireless communication systems, the propagation environment between
transmitter (Tx) and receiver (Rx) introduces complex distortions to the transmitted
signal. The baseband transmit signal is represented in complex form as:
\begin{equation}
    X = A e^{j\theta},
\end{equation}
where $A$ and $\theta$ denote the signal amplitude and phase.

When electromagnetic waves encounter obstacles in realistic environments,
they undergo reflection, diffraction, and scattering, giving rise to
multiple propagation paths. The composite received signal emerges as the
coherent superposition of these multipath components:
\begin{equation}
    Y = X \cdot \sum_{l=0}^{L-1} \alpha_l e^{j\phi_l},
\end{equation}
where $L$ is the number of distinct paths, each characterized by
attenuation $\alpha_l$ and phase shift $\phi_l$.

The Received Signal Strength Indicator (RSSI) summarizes the aggregate received power as a scalar measurement:
\begin{equation}
\mathrm{RSSI}
= 10 \log_{10}\!\left(||Y||^2\right/P_0),
\label{eq:rssi}
\end{equation}
where $P_0$ is the reference power. It reflects the combined effect of path loss, shadowing, and multipath
fading. 

The spatial distribution of received power is described by the angle--power spectrum $\Psi(\alpha, \beta)$, which quantifies the relative power arriving from azimuth angle $\alpha$ and elevation angle $\beta$.
$\Psi(\alpha, \beta)$ is obtained by evaluating the beam-steered relative power directly on a dense discrete angular grid:
\begin{equation}
    \Psi(\alpha, \beta) =
    \left|\mathbf{a}^{H}(\alpha, \beta)\,\mathbf{y}\right|^2,
\end{equation}
where $\mathbf{a}(\alpha, \beta)$ denotes the array steering vector and
$\mathbf{y}$ is the received signal vector. The angular grid is typically
sampled at $1^\circ$ resolution, yielding $N = 360 \times 90$ discrete
points. The finite antenna aperture limits the achievable angular
resolution, so the spectrum smoothness is governed by the array response
and grid sampling.

\subsection{Geometric Algebra}

GA, rooted in Clifford’s framework, extends classical vector spaces into a single computational language that handles higher-dimensional geometric primitives and their transformations within one coherent calculus. Instead of switching among matrices for rotations and quaternions for orientations, GA offers a coordinate-free algebra in which transformations arise through multiplication, unifying representation and computation of geometry~\cite{dorst2022guided, Roelfs_2023}. 

\begin{figure*}[htbp]
\centering
\includegraphics[width=\linewidth]{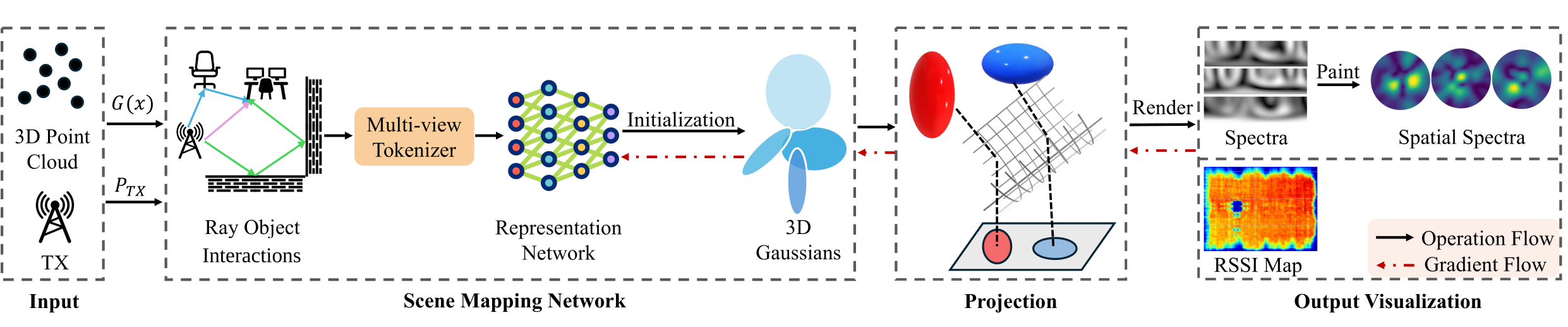}
\caption{GAI-GS structure. The tokenizer encodes interaction-aware representations from Gaussian primitives and transmitter context.}
\label{figure_framework}
\end{figure*}

We adopt the space-time algebra $\mathcal{G}_{3,0,1}$ with three spatial and one temporal dimension. Its elements form a graded structure: grade 0 scalars, grade 1 vectors, grade 2 bivectors, grade 3 trivectors, and grade 4 pseudoscalars. Each grade corresponds to a geometric entity such as points, lines, planes, volumes, and oriented hypervolumes, and the algebraic degree aligns with geometric dimensionality.

The central operation is the geometric product of vectors $\mathbf{v}_1$ and $\mathbf{v}_2$:
\begin{equation}
\mathbf{v}_1\mathbf{v}_2 = \mathbf{v}_1 \cdot \mathbf{v}_2 + \mathbf{v}_1 \wedge \mathbf{v}_2,
\end{equation}
which splits into a symmetric inner product that yields scalar projection and an antisymmetric wedge product that produces an oriented plane. We use an orthonormal basis ${\mathbf{e}_1,\mathbf{e}_2,\mathbf{e}_3,\mathbf{e}_4}$ in which the spatial basis elements square to $+1$ and the temporal element squares to $-1$, giving the Minkowski signature and allowing Lorentz boosts and spatial rotations to be expressed without matrix exponentials. Recent neural architectures that incorporate GA report stronger equivariance and improved representational efficiency, reinforcing its suitability as a backbone for learning geometric transformations~\cite{10.5555/3618408.3619627, brehmer2023geometric, NEURIPS2024_277628cf}. 

\section{Related Work}
\label{sec:Related}
\paragraph{3D-GS in Wireless Fields.}
A growing line of work adapts 3D-GS from vision to wireless frequency (RF), treating the wireless environment as a radiance field that can be learned from sparse measurements and rendered at arbitrary transceiver poses. RF-3DGS reconstructs a radio radiance field and renders spatial spectra within milliseconds after training, further exposing spatial-channel state information (CSI) of dominant paths from sparse samples, demonstrating site-specific channel modeling advantages over empirical and ray-tracing baselines~\cite{zhang2025rf3dgswirelesschannelmodeling}. Building on this direction, WRF-GS formulates wireless radiation field reconstruction with 3D-GS and introduces a physics-augmented variant WRF-GS+ that improves RSSI and CSI prediction while preserving real-time synthesis, highlighting the benefit of coupling electromagnetic (EM) priors with explicit Gaussian primitives~\cite{11044513}. Finally, GSRF extends 3D-GS to complex-valued RF fields via a Fourier–Legendre basis and RF-customized CUDA kernels, synthesizing RSSI, spatial spectra, and complex CSI with markedly lower training and inference cost than NeRF-style baselines~\cite{yang2025gsrf}.

\section{Framework}
\label{sec:framework}

\subsection{Overview}
We introduce GAI-GS, a geometric algebra–informed Gaussian-splatting framework for wireless channel modeling in Fig.~\ref{figure_framework}. The method injects explicit 3D scene geometry and device poses into the learning pipeline so the network internalizes ray–object interactions rather than treating them as black-box correlations. Our framework consists of two main components: (i) a \emph{Scene Mapping Network}, (ii) a \emph{Projection and Render Module}. Initially, the wireless measurements and the initialized 3D point cloud are sent into the Scene Mapping Network~\cite{park2019deepsdf} to represent the virtual transmitters for a set of 3D Gaussians, along with the attenuation and signal properties. Next, the Projection module projects the virtual transmitters onto the RX antenna plane using the Mercator projection, and then renders the projected 2D Gaussians under EM propagation constraints, aggregating distributed interactions into a unified spatial--frequency representation. 

Specifically, inspired by the classification tokens used in large language and multimodal encoders~\cite{47751, pmlr-v162-li22n, pmlr-v202-li23q}, we design a multi-view tokenizer that implicitly converts a set of ray-object interactions into a global scene token. The global token aggregates scene-level context from Gaussian primitives. This tokenizer implicitly learns ray--object interactions and captures propagation characteristics in the wireless environment. 

Elementary interactions are parameterized with rotors in geometric algebra, denoted $R_r$, $D$, $T$, and $R_0$ for reflection, diffraction-like bending, transmission, or refraction, respectively, and initial alignment to the scene frame. Each rotor acts on a ray state vector through a sandwich product, and complex paths arise by composition. Let $\mathbf{V}$ encode the state of the ray, such as its direction, wave vector, or signal attributes. The cumulative effect of multiple interactions is
\begin{equation}
\mathbf{V}' = \boldsymbol{I}\mathbf{V}\boldsymbol{I}^{-1},
\end{equation}
where the versor $\boldsymbol{I}$ is the learned product of the relevant rotors selected by attention. The rotors are produced implicitly by the geometric encoder from the multi-view tokens and are optimized end-to-end under wireless supervision. This unified algebraic representation captures multi-bounce, multi-effect propagation within a single differentiable mechanism and removes the need for separate specialized modules for reflection, refraction, and diffraction.

\subsection{Multi-view Tokenizer}

\paragraph{Geometric Algebra Transformer.}


We adopt the geometric algebra Transformer (GATr)~\cite{brehmer2023geometric} as the encoder to extract global embeddings before feeding data into our model. Within the geometric algebra space $\mathbb{G}_{3,0,1}$, geometric transformations such as rotations, reflections, diffractions, and transmissions can be compactly expressed using sandwich products, i.e., $V' = IVI^{-1}$ where $I$ is a multivector-valued interaction operator. This algebraic framework provides a compact and physically consistent way to represent ray–object interactions in wireless propagation and allows us to encode these interactions directly in the feature space rather than relying on explicit geometric annotations.

In realistic wireless environments, ray–object interactions are highly complex: a single received ray often results from multiple reflections, edge diffractions, and penetrations through heterogeneous materials, with interactions occurring at unknown surface locations and orientations. Classical models, such as ray tracing, require detailed information about scene geometry, material properties, and precise collision points in order to explicitly construct each interaction operator. This dependency makes large-scale modeling cumbersome and most neural wireless models therefore either ignore explicit ray–object structures or approximate them with hand-crafted features. In contrast, our approach is the first to implicitly model in-scene ray–object interactions, using learned geometric algebra operators to guide wireless scene representation learning without requiring material labels or explicit interaction locations.

In wireless environments, EM rays interact with surfaces following geometric algebraic rules. For a surface reflection, the incident ray $\mathbf{x}$ and surface normal $\hat{n}$ yield the reflected ray via a sandwich product $\mathbf{x}' = -R\mathbf{x}R^{-1}$, where $R$ encodes the rotation associated with the reflection and corresponds to the operators $R_r$ and $R_0$. 
Edge diffraction can be represented analogously as $\mathbf{x}' \approx D\mathbf{x}D^{-1}$ using a diffraction operator $D$, where the approximation arises because diffraction alters amplitude, phase, and spatial energy distribution in a nonlinear manner that cannot be captured by a purely geometric mapping alone. Material penetration can be expressed deterministically as $\mathbf{x}' = T\mathbf{x}T^{-1}$ with $T$ encoding the transmission effect of a given material interface. A full ray path with $n$ successive interactions, reflections, diffractions, and transmissions, can be represented as a sequential composition:
\begin{equation}
V' = I_1 I_2 \cdots I_n V I_n^{-1} \cdots I_2^{-1} I_1^{-1} = I V I^{-1},
\end{equation}
where each $I_i$ denotes an individual ray–object interaction and $I = \prod_{i=1}^n I_i$ is the aggregate interaction operator for that path. This formulation indicates that a physically consistent ray trajectory is completely determined by its cumulative geometric algebra operator $I$, which we learn implicitly from data.

The geometric algebra attention mechanism used in GATr exhibits a structural correspondence with these physical transformations. Let $q$, $k$, and $v$ denote input tensors with $n_c$ channels. Standard dot-product attention aggregates information as:
\begin{equation}
\text{Attention}(q, k, v)_{i'c'} = \sum_i \text{Softmax}_i \left( \frac{\langle q_{i'c}, k_{ic} \rangle}{\sqrt{8n_c}} \right) v_{ic'},
\end{equation}
which can be interpreted in a sandwich-product form:
\begin{equation}
\text{Attention}(q, k, v)_{i'c'} = \sum_i A_{i'} v_{ic'} A_{i'}^{-1},
\end{equation}
where $A_{i'}$ is a multivector-valued operator constructed from the attention weights for query index $i'$, the indices $i$ and $i'$ denote tokens, and $c$ and $c'$ index channels. Under this view, each $A_{i'}$ acts as a learned interaction operator, analogous to $I$ above, that transforms value features $v_{ic'}$ into a representation consistent with the aggregate effect of all paths contributing to token $i'$.

By embedding geometric algebra directly into the attention computation, GATr enforces rotational and reflectional equivariance and aligns the learned feature space with fundamental EM propagation symmetries. As a result, the encoder can implicitly infer complex, multi-bounce ray–object interactions from data, without explicit knowledge of material types or collision locations.

\paragraph{Multi-view Tokenizer.}
We first use the GATr to extract tokens that encode ray–object interaction patterns in the scene, capturing how rays are rotated, reflected, diffracted, and attenuated as they propagate. 

A naive implementation would feed all $N$ Gaussian positions into GATr, which incurs quadratic attention cost when $N$ is large. To improve efficiency, we instantiate the tokenizer using a subset of Gaussians by selecting the top-M highest-opacity primitives as anchors, where $M \ll N$. These anchors typically correspond to geometrically salient regions such as walls, obstacles, and strong reflectors. Since the Gaussian representation evolves during training, the anchor subset is updated accordingly to remain consistent with the current scene representation. The \emph{CLS} output is broadcast to all $N$ Gaussians and serves as a global scene-level representation that aggregates dominant geometric and propagation context from the selected Gaussian anchors together with the transmitter-conditioned positional embeddings. In particular, it encodes the collective effect of ray--object interactions within the scene, including reflections, diffractions, and attenuation patterns, as captured by the geometric algebra attention. This shared representation provides a unified scene context that is subsequently used by both the attenuation and signal branches to ensure consistent propagation modeling across all Gaussian primitives. This reduces the complexity from $O(N^2)$ to $O(M^2)$, making the encoder agnostic to the total Gaussian count while preserving geometric expressiveness.

In parallel, we derive Euclidean position embeddings that preserve metric structure such as absolute locations, relative distances, and large-scale layout. Concatenating these two streams yields a unified multi-view embedding that combines interaction-aware features from geometric algebra with geometry-aware features from Euclidean space.
This complementary representation enables the network to distinguish scenes sharing similar transmitter configurations yet differing in intermediate interactions, and to resolve local interaction patterns that positions alone leave underconstrained.
The resulting embedding encourages the network to learn wireless scene representations anchored in both physical ray behavior and global spatial structure, improving data efficiency and robustness to layout changes.

\subsection{Mapping, Projection and Render Module}

\begin{figure}[htbp]
\centering
\includegraphics[width=\linewidth]{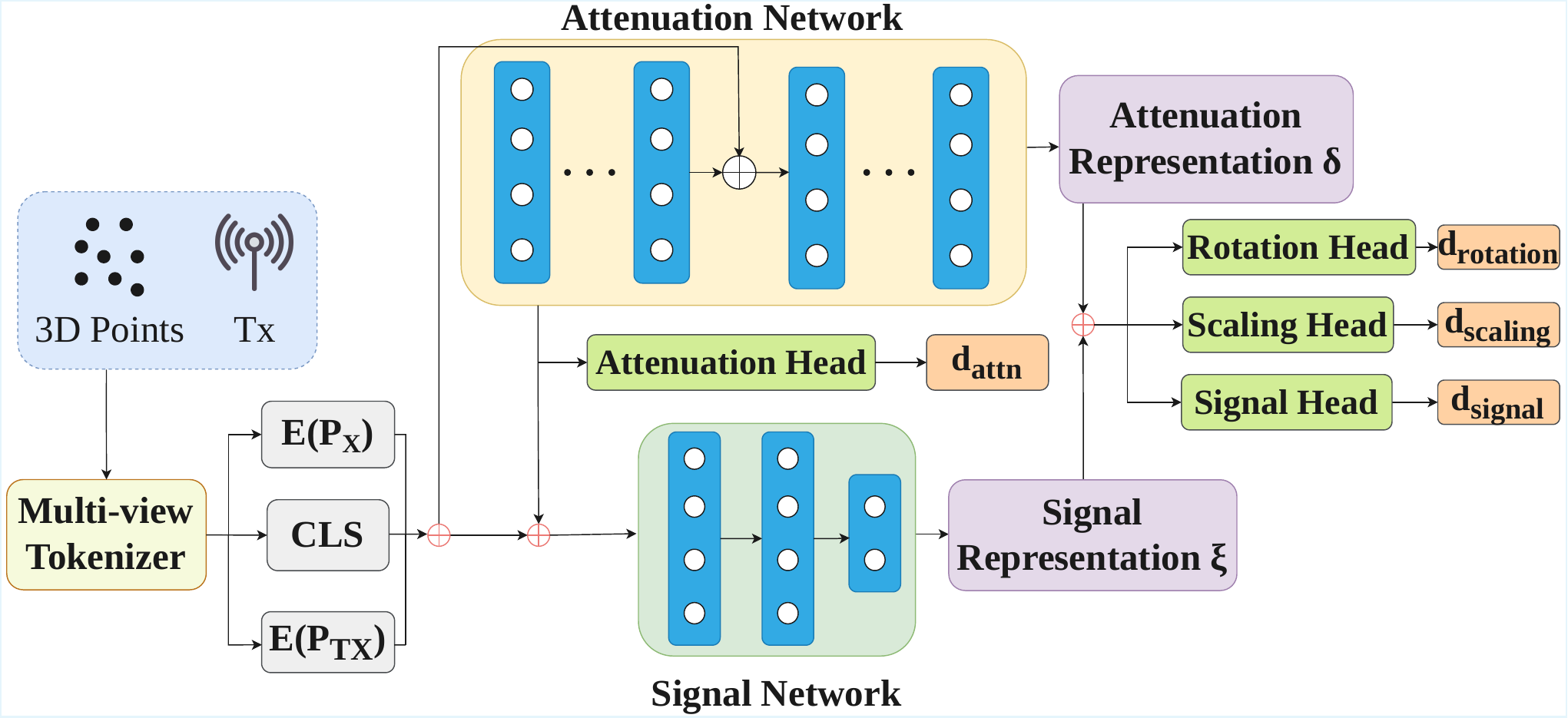}
\caption{Scene mapping model overview.}
\label{figure_model}
\end{figure}

\paragraph{Scene Mapping Network.}
Fig.~\ref{figure_model} depicts the structure of our scene mapping network. The representation network adopts a dual-network design with two complementary branches that decompose wireless propagation into distinct physical processes: attenuation and signal representations. The attenuation network predicts spatially varying decay coefficients that characterize the progressive weakening of EM waves as they traverse the scene. This branch embeds material- and medium-dependent effects and encodes the inherent attenuation properties of the volumetric environment, yielding a 3D field that maps position to expected signal extinction. In parallel, the signal network reconstructs the scattered-field distribution induced by interactions with scene geometry. It explicitly models non-line-of-sight propagation to capture signal characteristics.

Formally, $P_x \in \mathbb{R}^3$ denotes the center coordinate of the 3D Gaussian at location $x$, and $P_{TX} \in \mathbb{R}^3$ denotes the transmitter location. First, the multi-view tokenizer $F_{\text{mv}}$ encodes those inputs into a set of positional embeddings $E$ and a global scene token \emph{CLS}:
\begin{equation}
    CLS, E(P_\mathrm{x}), E(P_\mathrm{TX}) = F_{\text{mv}}(P_{\mathrm{x}}, P_{\mathrm{TX}}).
\end{equation}
Then the global token \emph{CLS} is concatenated to the positional embeddings of the $P_{\mathrm{TX}}$ and $P_{\mathrm{x}}$ to generate multi-view Tx tokens. Here, \emph{CLS} provides a shared scene-level context that captures the global interaction structure of the environment and guides both attenuation and signal prediction. 
Based on this representation, the attenuation network $F_{\text{att}}$ predicts a scalar attenuation field $\delta(\mathbf{x})$ and an intermediate geometric feature $f$ at each position:
\begin{equation}
    \delta(\mathbf{x}),\, f 
    = F_{\text{att}}( E(P_\mathrm{TX}), E(P_\mathrm{x}), CLS ).
\end{equation}

Subsequently, the signal network $F_{\text{sig}}$ predicts the signal strength conditioned on the intermediate feature, the \emph{CLS} token, and the embeddings of the Tx and Gaussian point:
\begin{equation}
    \xi(\mathbf{x})
    = F_{\text{sig}}\big(f, E(P_\mathrm{TX}), E(P_\mathrm{x}), \mathrm{CLS}\big),
\end{equation}
yielding a scattered-field amplitude that captures indirect paths such as reflections and diffuse scattering. Finally, we concatenate the attenuation feature and the signal representation
to form a joint feature vector, which is then passed through three dedicated MLP heads, namely a Rotation Head, a Scaling Head, and a Signal Head. The Rotation and Scaling Heads produce residual updates, $d_{\text{rotation}}$ and $d_{\text{scaling}}$, which are applied to the original Gaussian rotation and scaling parameters, respectively, allowing the network to model geometric deformations in a residual manner rather than regressing absolute values. For the signal branch, the Signal Head operates in the spherical harmonics (SH) coefficient space, producing a residual $d_{\text{signal}}$ that is added to the original SH coefficients of each Gaussian to obtain the updated signal representation: $\tilde{\xi}(x_i)=\xi(x_i)+d_{\text{signal},i}$. This residual parameterization ensures that the network learns deviations from the canonical Gaussian parameters, which facilitates stable training and preserves the structural priors encoded in the original 3D Gaussians.

In standard 3D Gaussian Splatting, each Gaussian primitive carries a fixed opacity $\alpha_i$ that is invariant to the query condition. However, in wireless propagation, the effective attenuation of a spatial region is transmitter-dependent: an obstacle may fully occlude the signal from one transmitter while remaining largely transparent to another. To account for transmitter-dependent attenuation, we use the attenuation intermediate feature to parameterize an opacity adjustment for each Gaussian. Concretely, the effective opacity is written as $\tilde{\alpha}_i = \alpha_i + d_{\text{attn}, i}$, where $d_{\text{attn}, i}$ is a learned residual conditioned on the geometric feature $f$. An $L_2$ penalty on $d_{\text{attn}}$ regularizes this adjustment.

We implement both the attenuation and signal networks with MLPs. For the attenuation network, we further incorporate residual connections to stabilize training and preserve geometric information.

\begin{figure}[!htbp]
  \centering
  \begin{subfigure}[t]{0.3\linewidth}
    \centering
    \includegraphics[width=\linewidth]{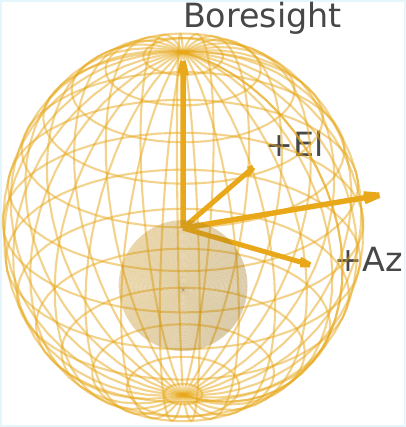}
    \caption{3D Gaussian on AoA sphere.}
    \label{fig:AoA}
  \end{subfigure}
  \hfill
  \begin{subfigure}[t]{0.65\linewidth}
    \centering
    \includegraphics[width=\linewidth]{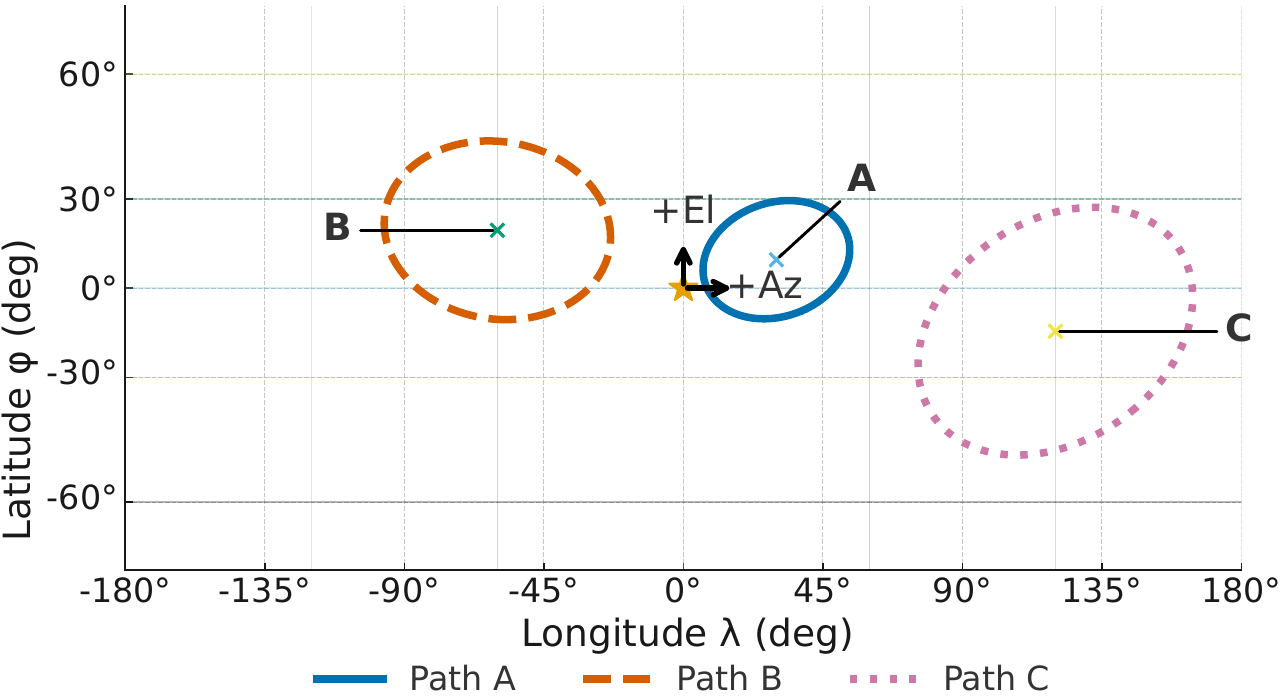}
    \caption{Mercator projection of angular Gaussians on RX perception plane.}
    \label{fig:Mercator}
  \end{subfigure}
  \caption{Mercator projection module, where $E_l$, $A_z$, and AoA denote elevation, azimuth, and angle of arrival, respectively.}
  \label{fig:Projection}
\end{figure}

\paragraph{Projection and Render Module.}

We follow the projection and rendering principles introduced in WRF-GS~\cite{11044513} and WRF-GS+~\cite{wen2024neural} to process the 3D Gaussians. Specifically, the 3D Gaussian representations are projected onto the perception plane of the RX antenna array using the Mercator projection. As shown in Fig.~\ref{fig:AoA}, the spherical coordinate system captures the antenna's field of view through azimuth and elevation angles, denoted as $Az$ and $El$ respectively, which are measured relative to the boresight direction. The shaded region represents the antenna's effective coverage area containing the Gaussian distributions.

Fig.~\ref{fig:Mercator} illustrates the Mercator projection used to map 3D Gaussian lobes onto the RX perception plane. Longitude $\lambda$ is mapped linearly to the horizontal axis in $[-180^\circ, 180^\circ]$, while latitude $\phi$ is mapped nonlinearly to vertical coordinates in $[-60^\circ, 60^\circ]$, preserving local angles and shapes and thus antenna directivity. Formally, the Mercator mapping from spherical angles $(\lambda, \phi)$ to 2D perception-plane coordinates $(u, v)$ is given by $u = \lambda$ and $\qquad 
v = \alpha \log \tan\!\left(\frac{\pi}{4} + \frac{\phi}{2}\right),$
where $u$ and $v$ denote the horizontal and vertical coordinates on the RX perception plane, respectively, and $\alpha$ is a scaling factor selected such that $v \in [-60^\circ, 60^\circ]$. Three representative paths highlight the transformation: Path A (blue ellipse) shows that an off-center Gaussian lobe preserves its elliptical structure; Path B (orange dashed circle) depicts a symmetric distribution at an off-center position; and Path C (pink dotted circle) illustrates how Gaussians are mapped at different angular positions. This angle-preserving mapping allows 3D Gaussian primitives to retain their geometric relationships in the 2D perception plane, enabling faithful representation of spatial signal characteristics.

Then, projected 2D Gaussian primitives are binned into tiles for massively parallel processing. Each tile operates independently on the primitives that intersect its spatial extent; primitives spanning multiple tiles are replicated so each tile has a complete local set. Within a tile, primitives are sorted in depth to preserve physically consistent signal accumulation.

In the RF setting, each projected Gaussian primitive acts as a virtual transmitter characterized by an updated signal representation $\tilde{\xi}(x_i)$ and contributes to the attenuation field through its effective opacity. Specifically, the contribution of the $i$-th primitive at angular coordinate $x$ is defined as: 
\begin{equation}
    S_i(x)=\tilde{\xi}(x_i)\prod_{j=1}^{i-1}\delta(x_j).
\end{equation}

The final received signal at pixel $k$ is obtained by accumulating the contributions from all projected virtual transmitters:
\begin{equation}
   R_k=\sum_{i=1}^{N} S_i(x_i)\,\tilde{\alpha}_i \prod_{j=1}^{i-1}(1-\tilde{\alpha}_j), 
\end{equation}
where $\tilde{\alpha}_i=\alpha_i+d_{\text{attn},i}$ denotes the effective opacity of the $i$-th Gaussian.

The resulting spatial power map approximates the wireless radiation field and provides actionable channel information for system design and optimization. Additionally, for RSSI map construction, we standardize the RSSI map by subtracting its mean, scaling with a temperature threshold, and stabilizing by shifting with the map’s maximum; a softmax over all pixels then produces attention weights blended with a uniform prior to ensure coverage. The final scalar RSSI map is obtained as the attention-weighted average of the map, which emphasizes strong-signal regions while remaining robust to noise and outliers.

\begin{figure*}[htbp]
  \centering
  \begin{tabular}{@{}c@{\hspace{\colgap}}c@{\hspace{\colgap}}c@{\hspace{\colgap}}c@{\hspace{\colgap}}c@{\hspace{\colgap}}c@{\hspace{\colgap}}c@{\hspace{\colgap}}c@{}}
    \colstack{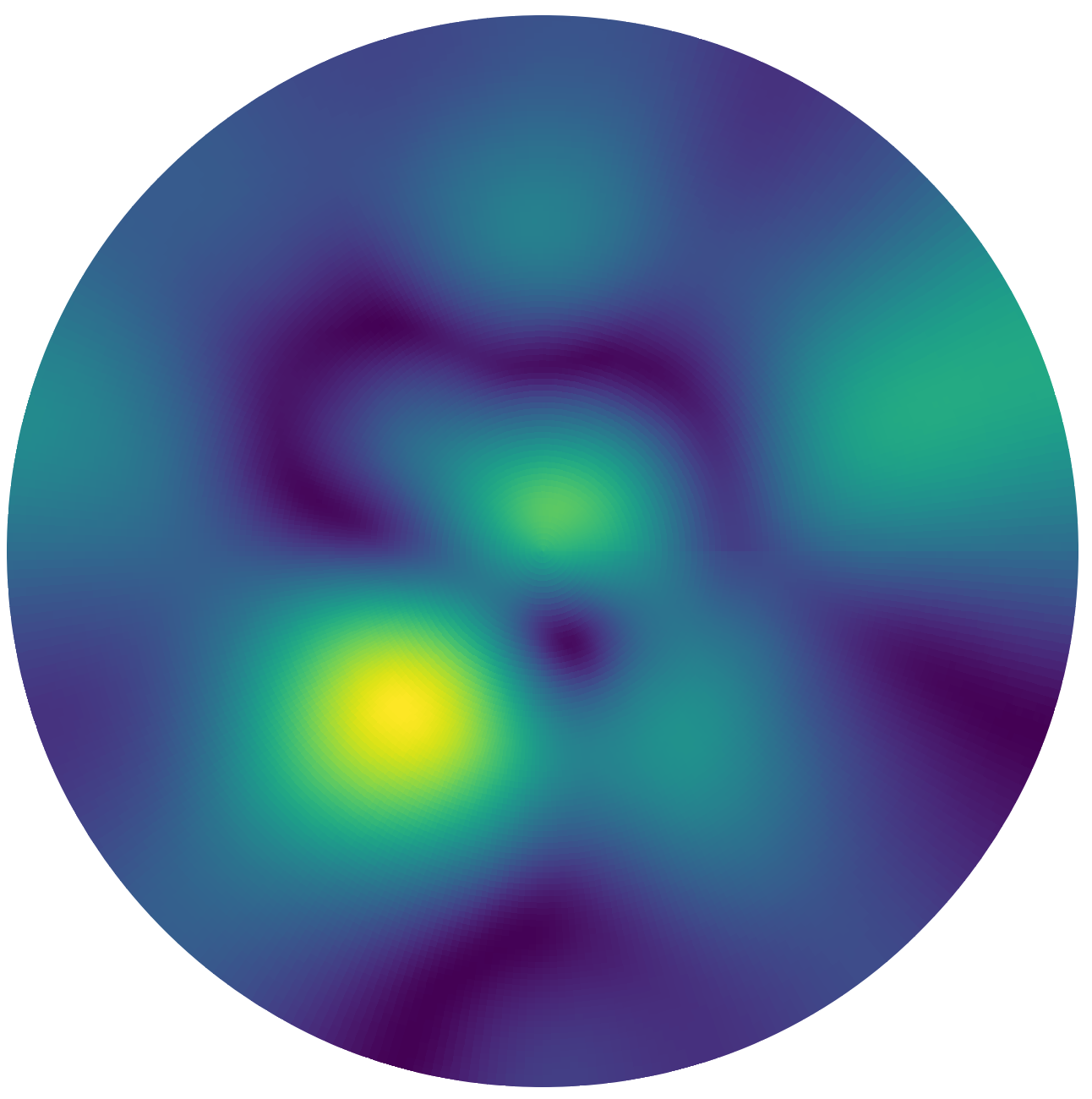}{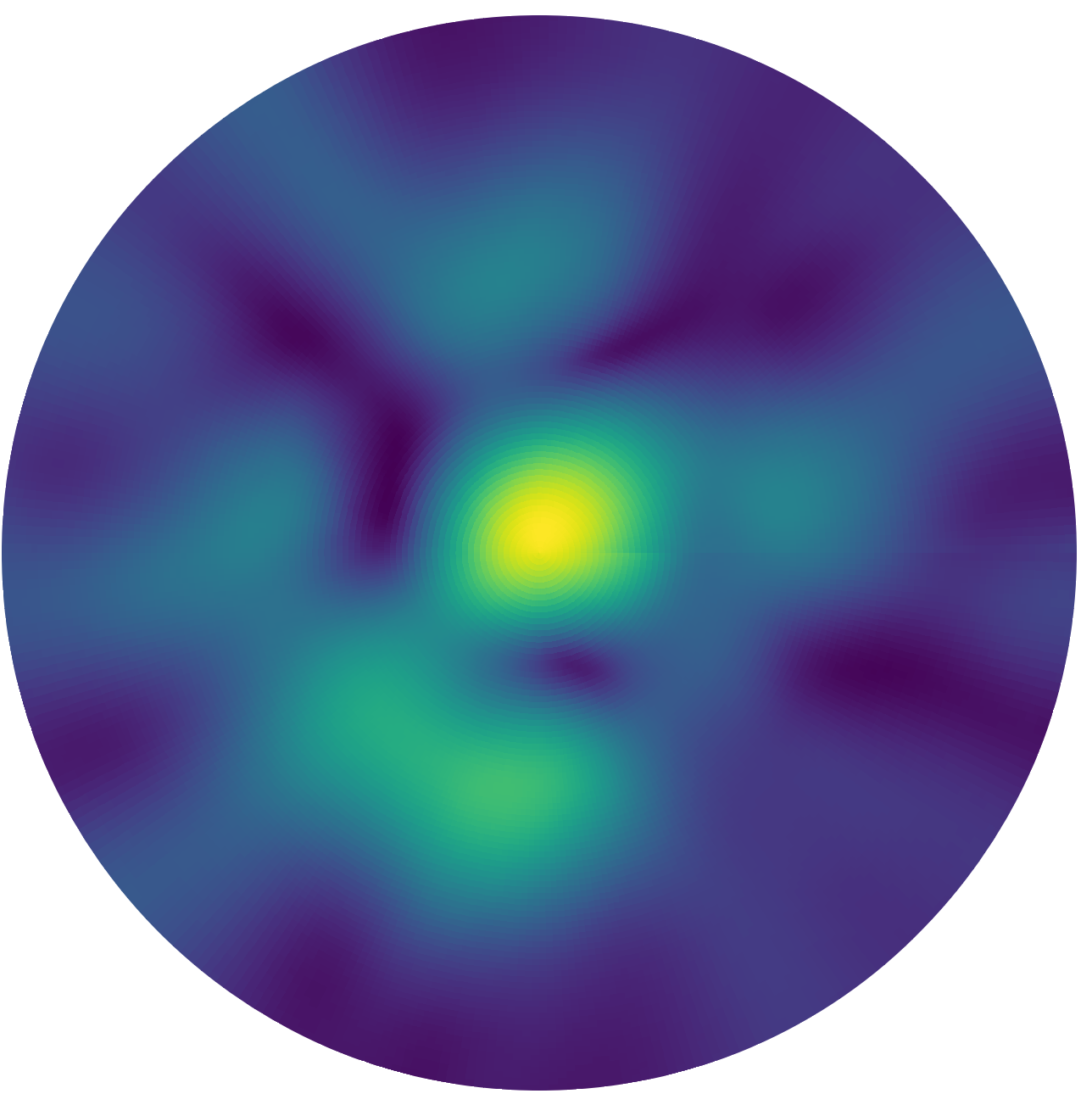}{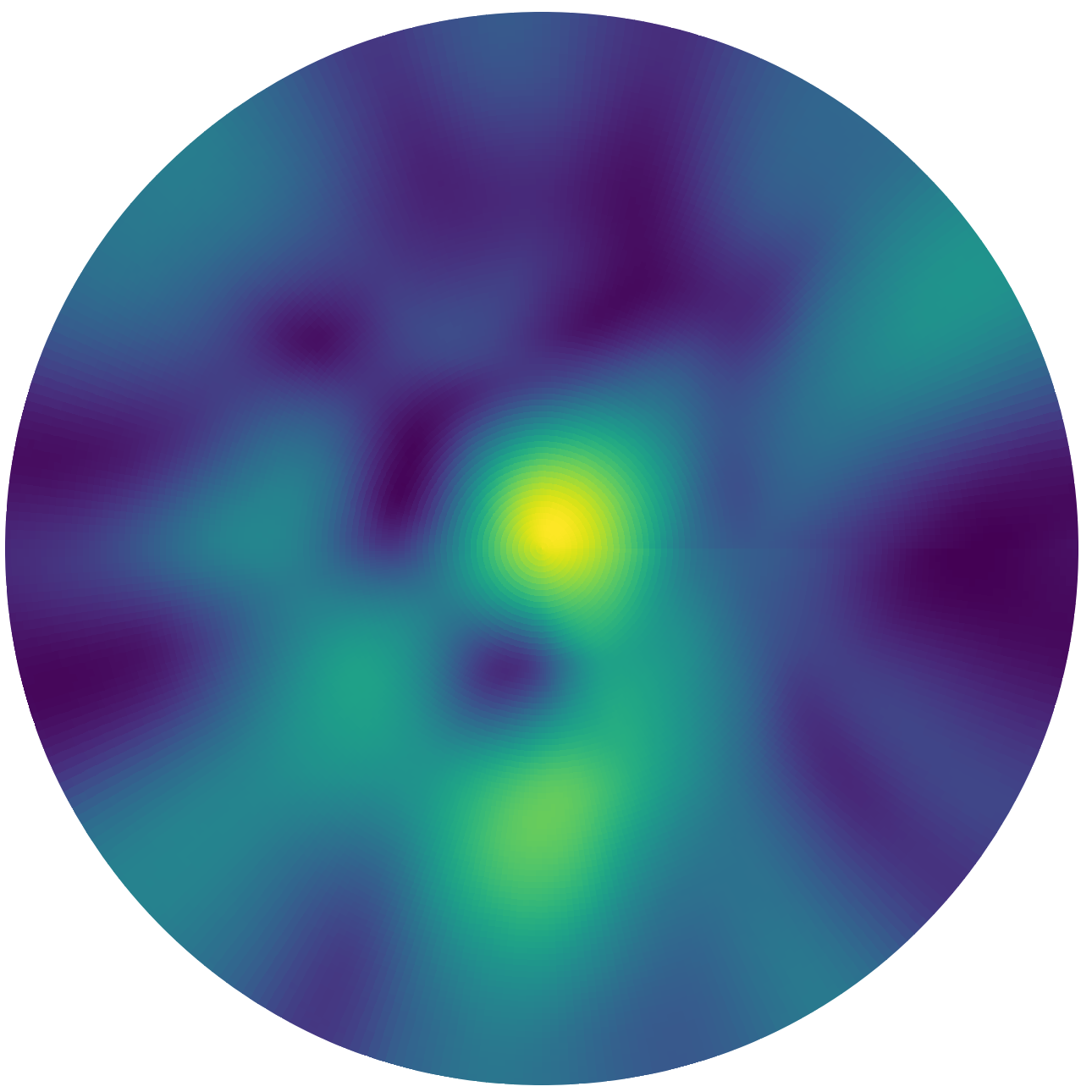} &
    \colstack{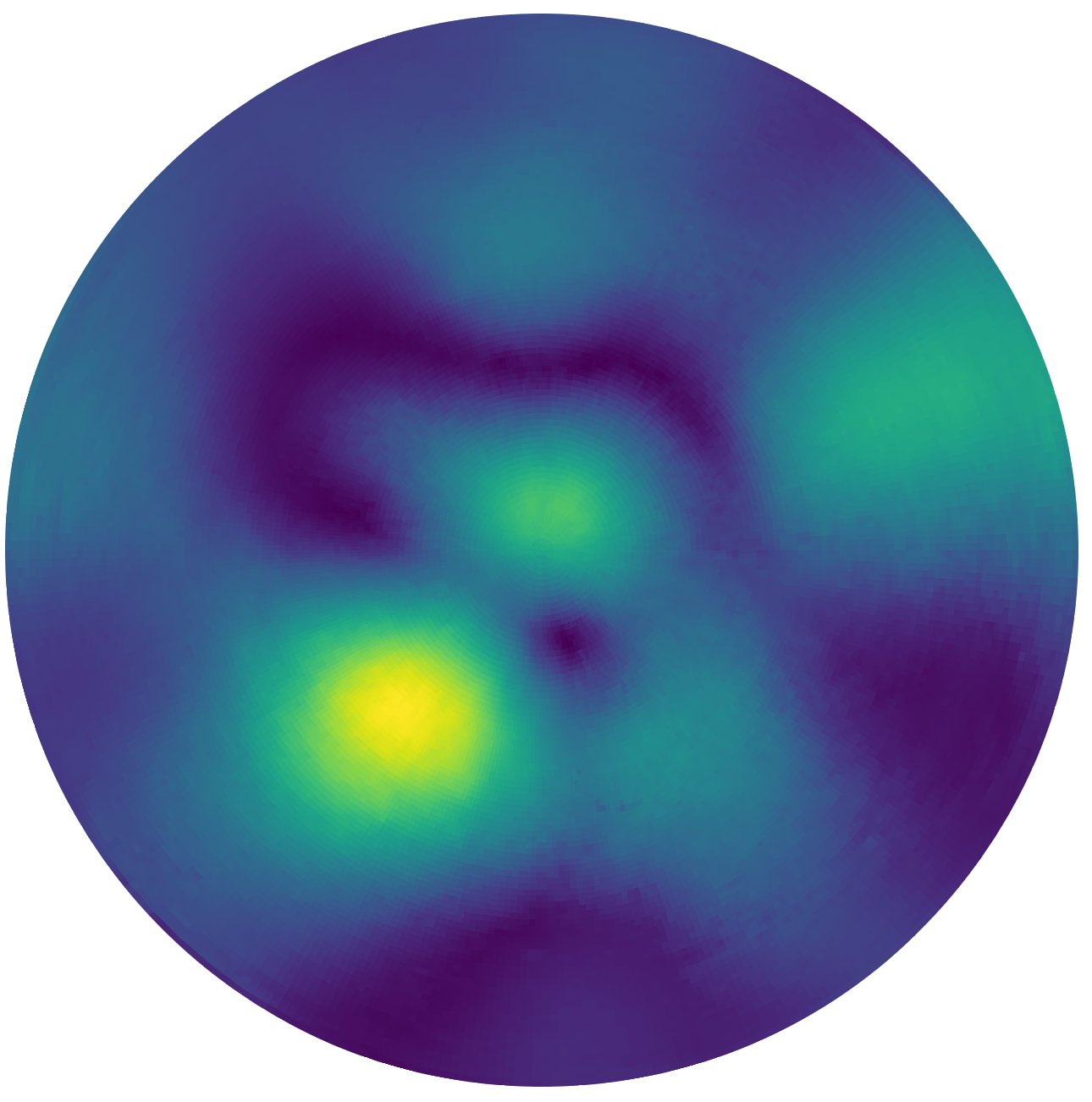}{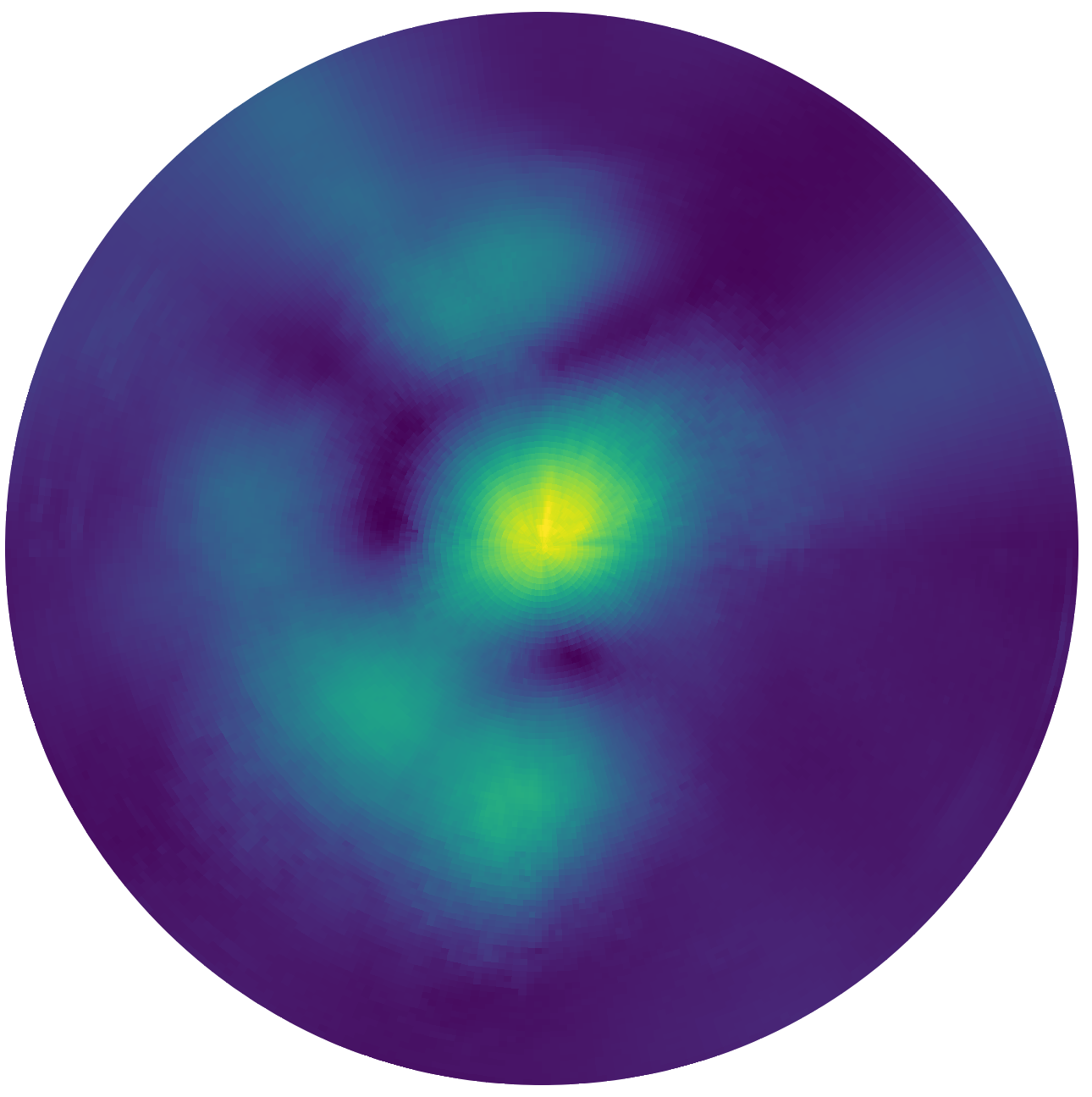}{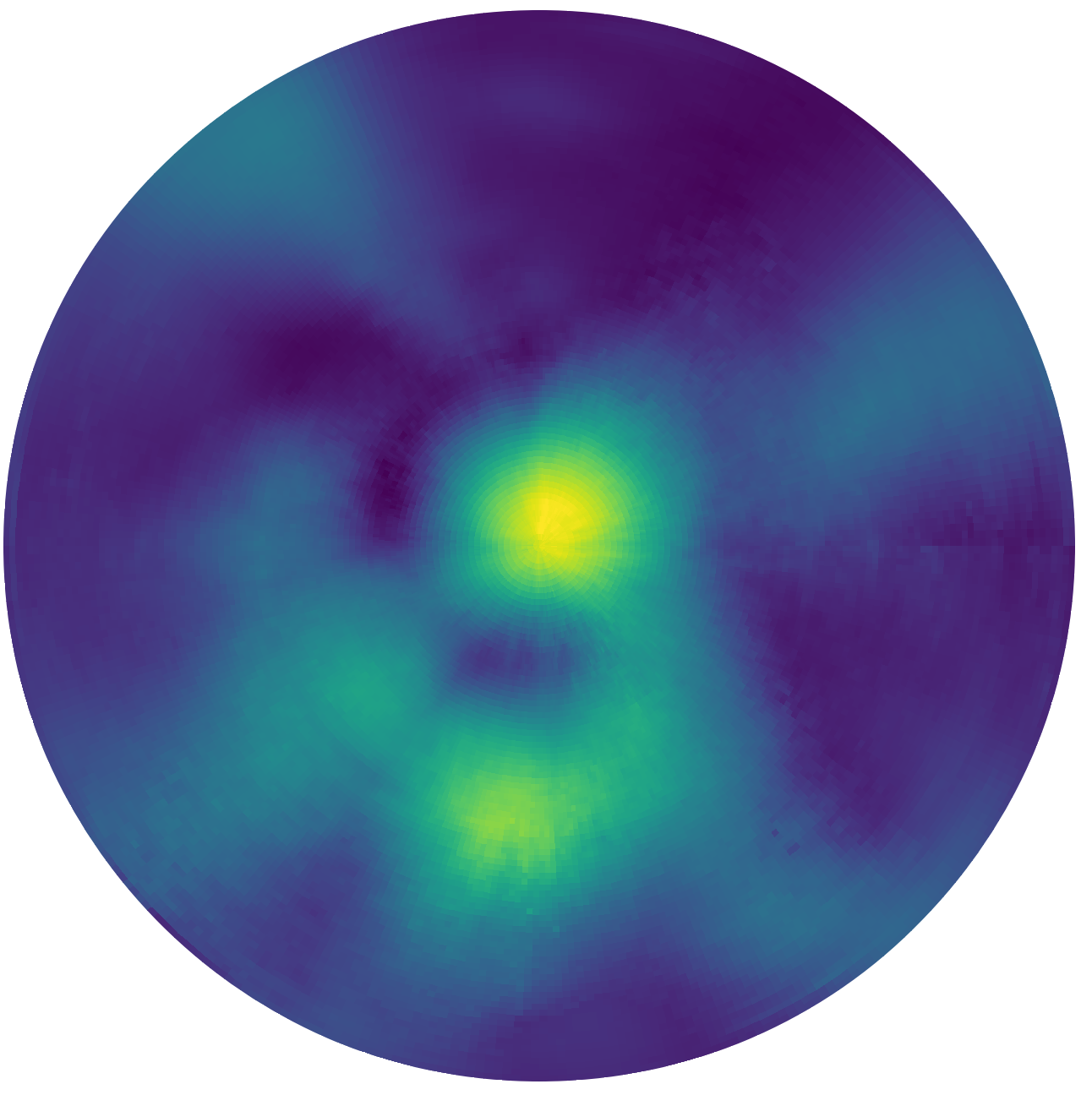} &
    \colstack{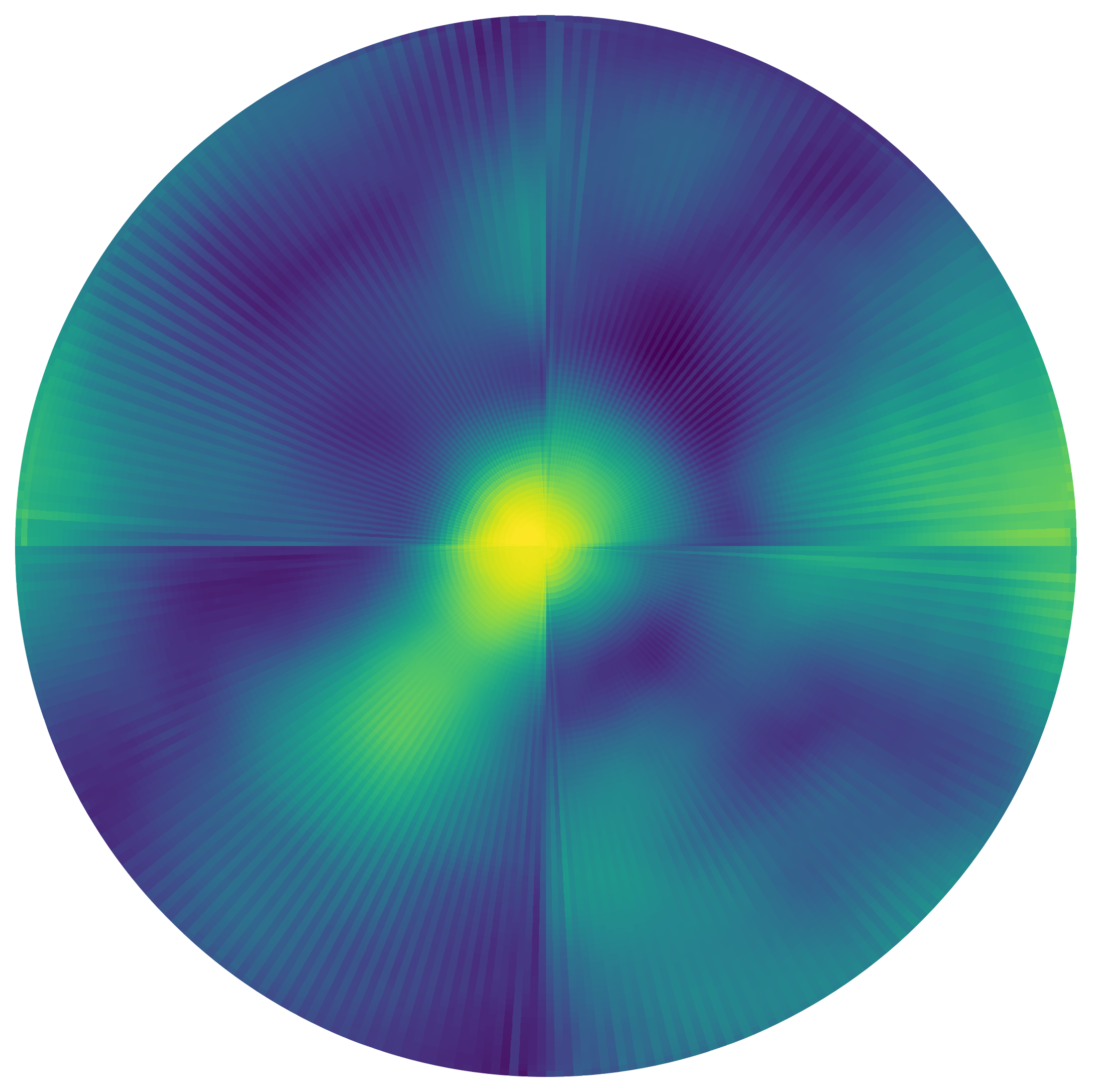}{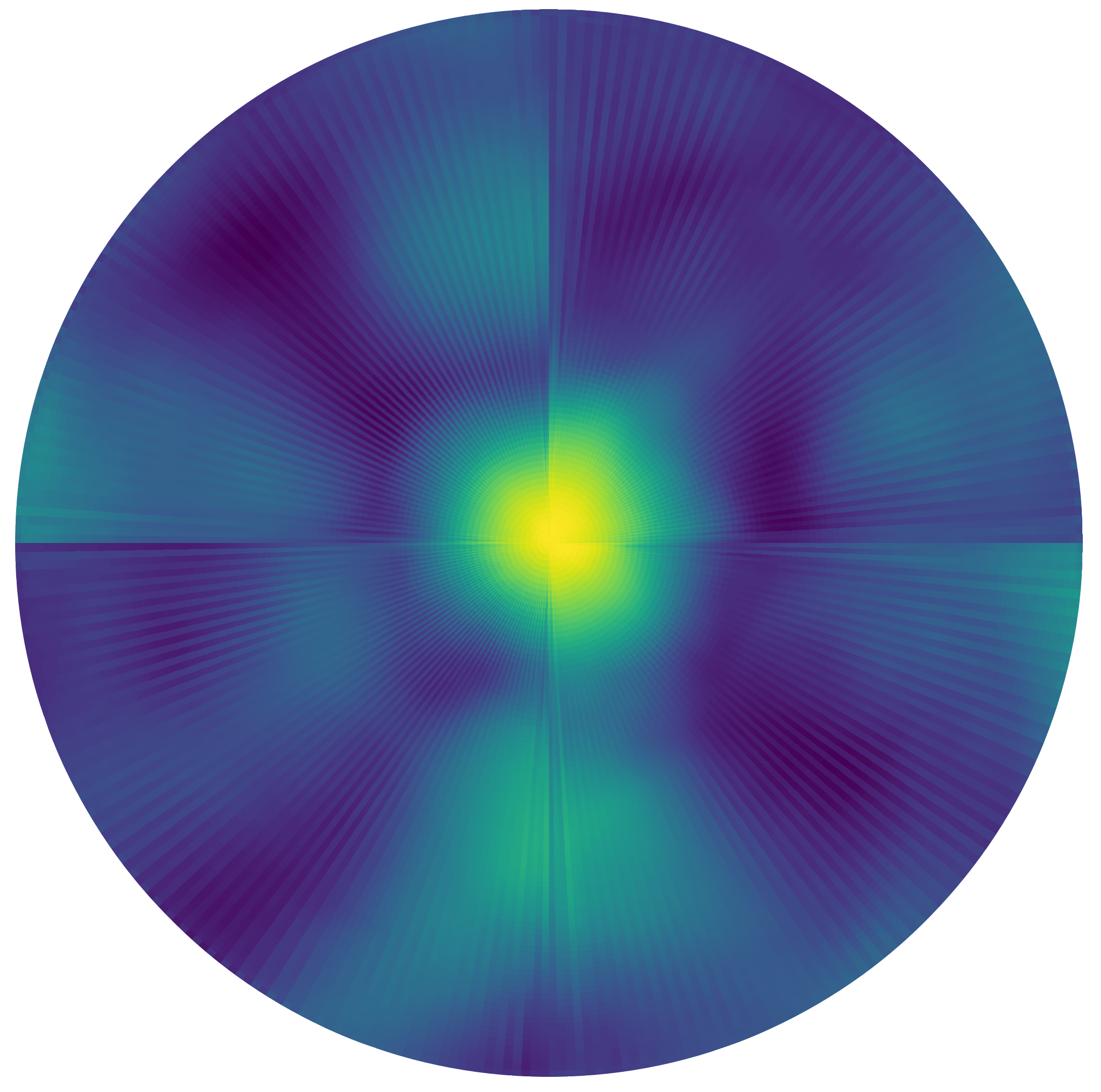}{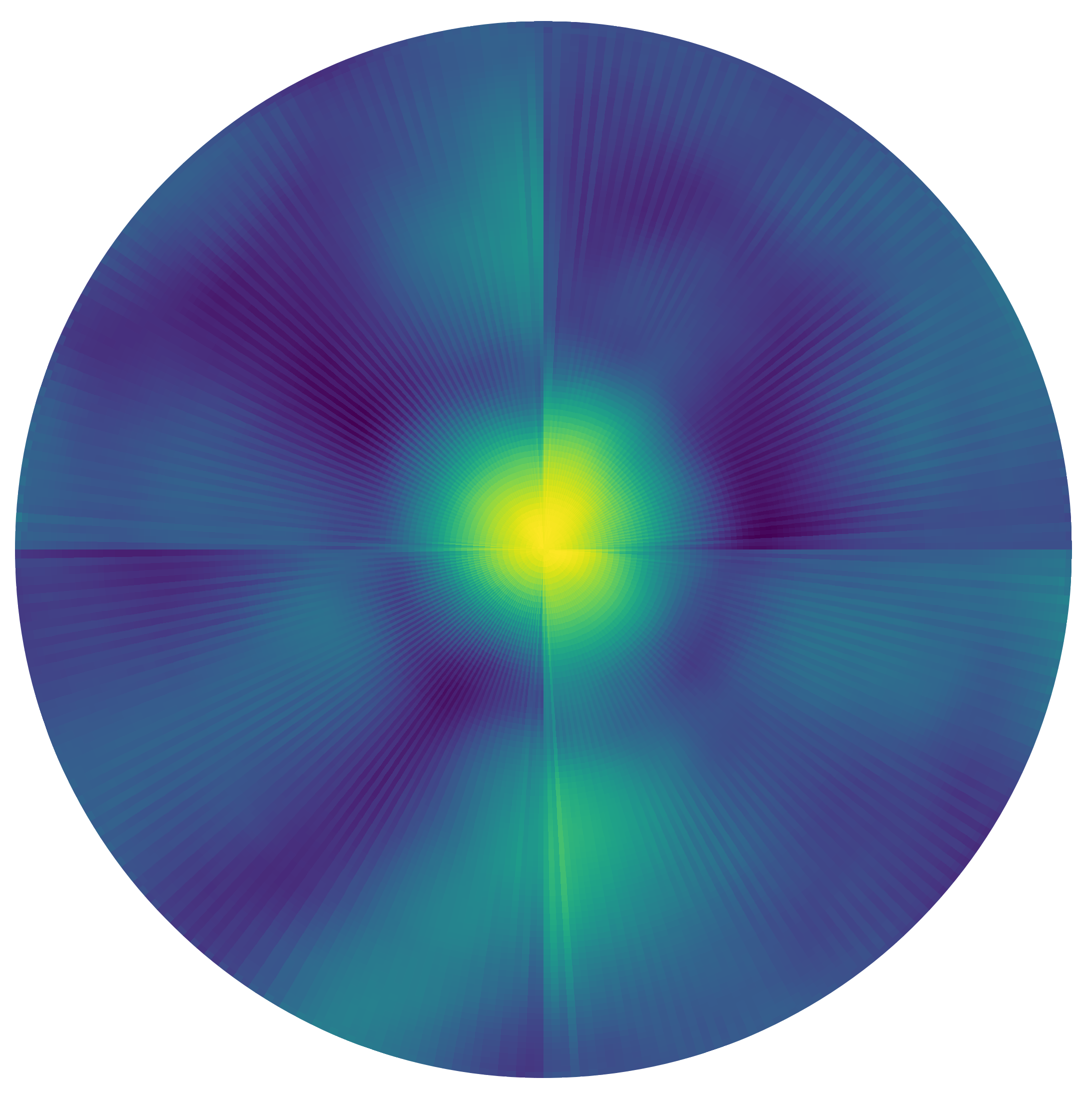} &
    \colstack{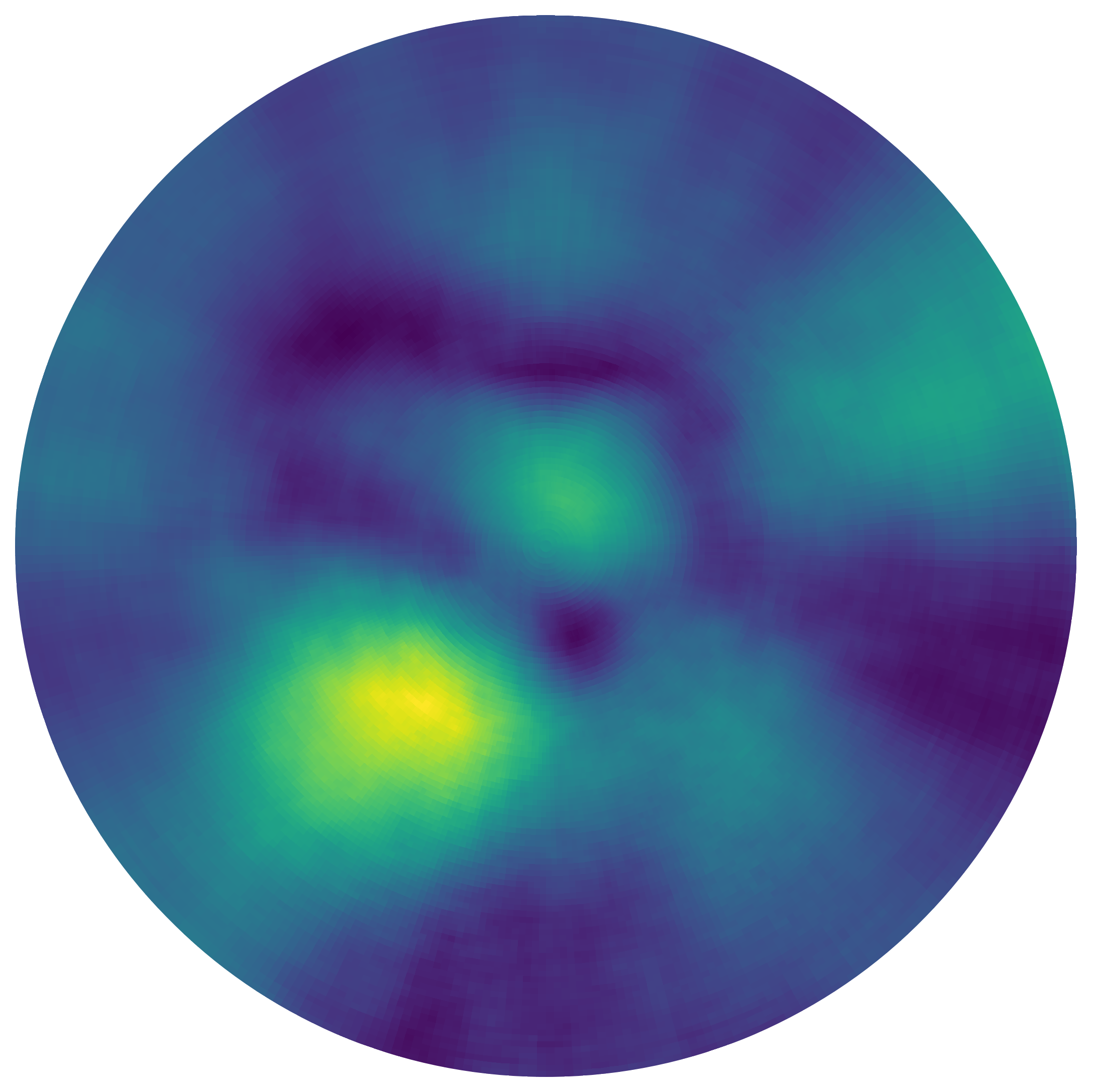}{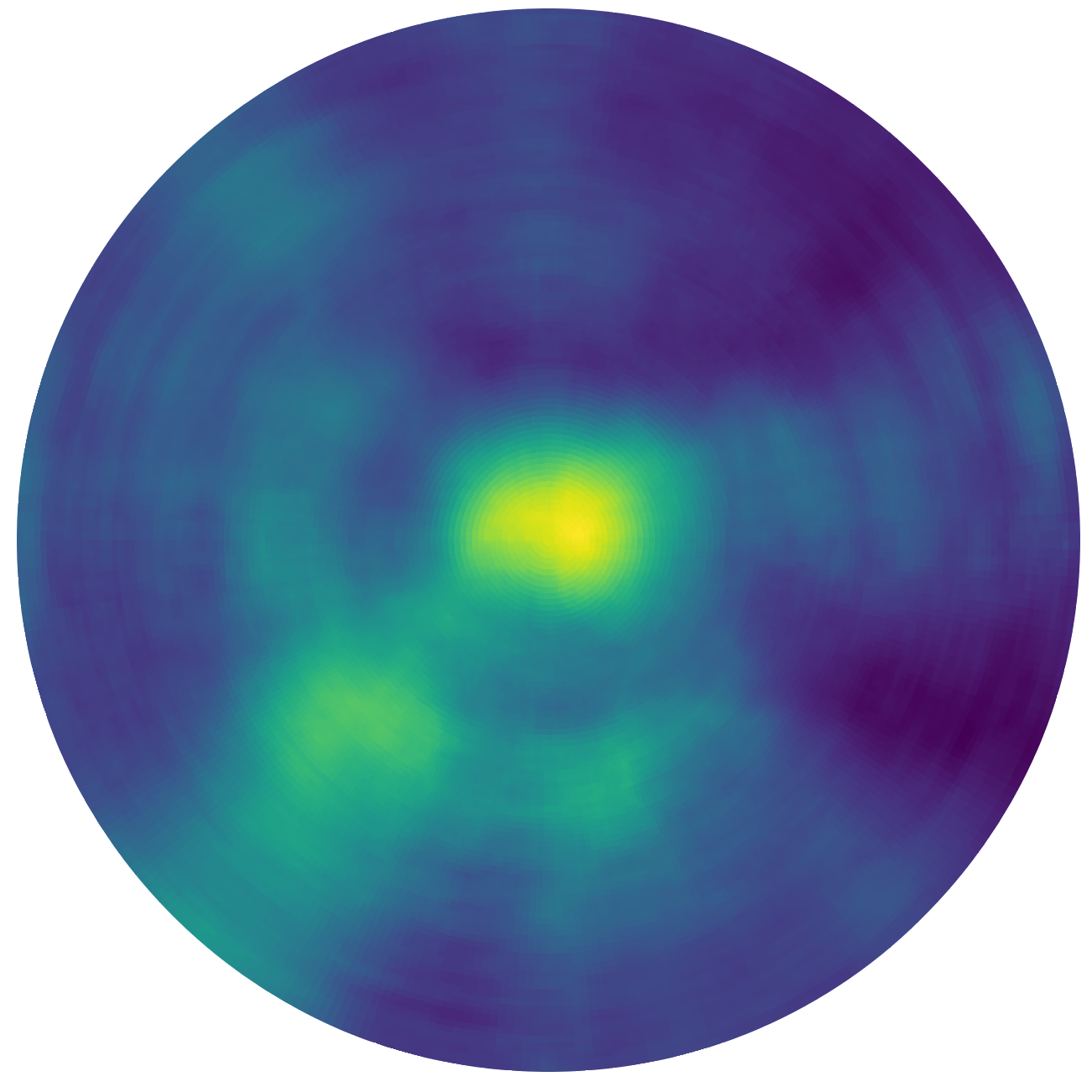}{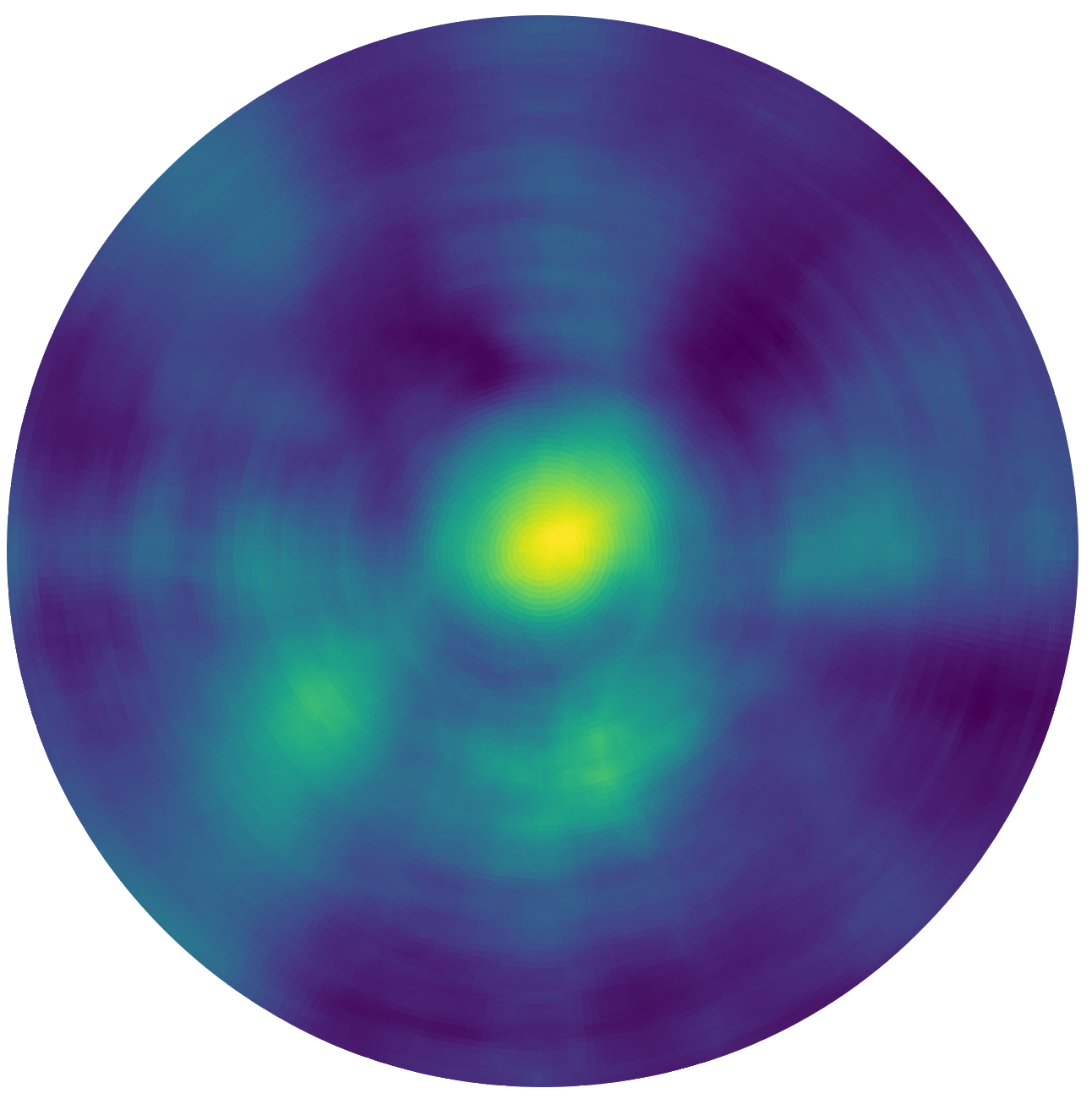} &
    \colstack{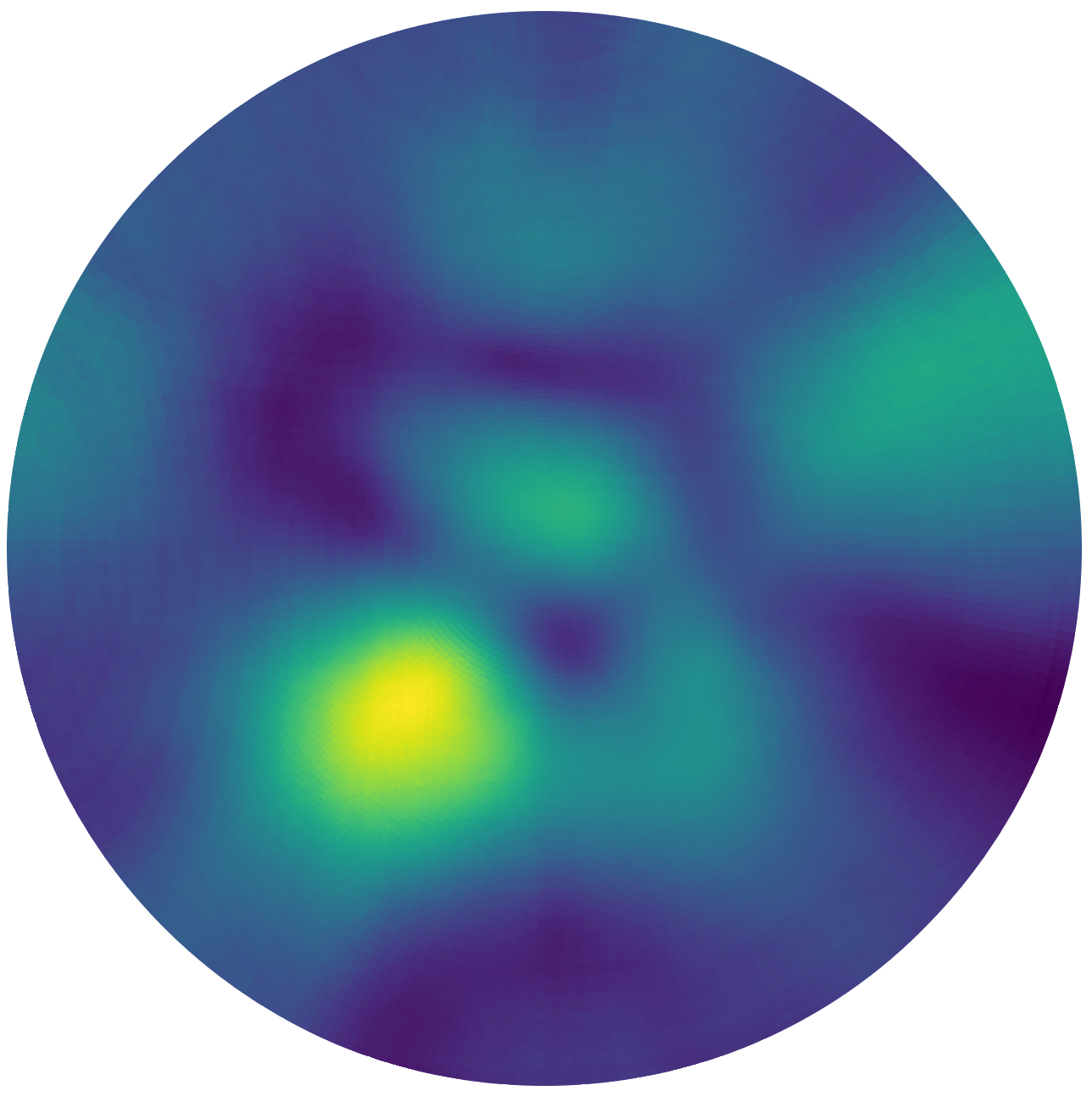}{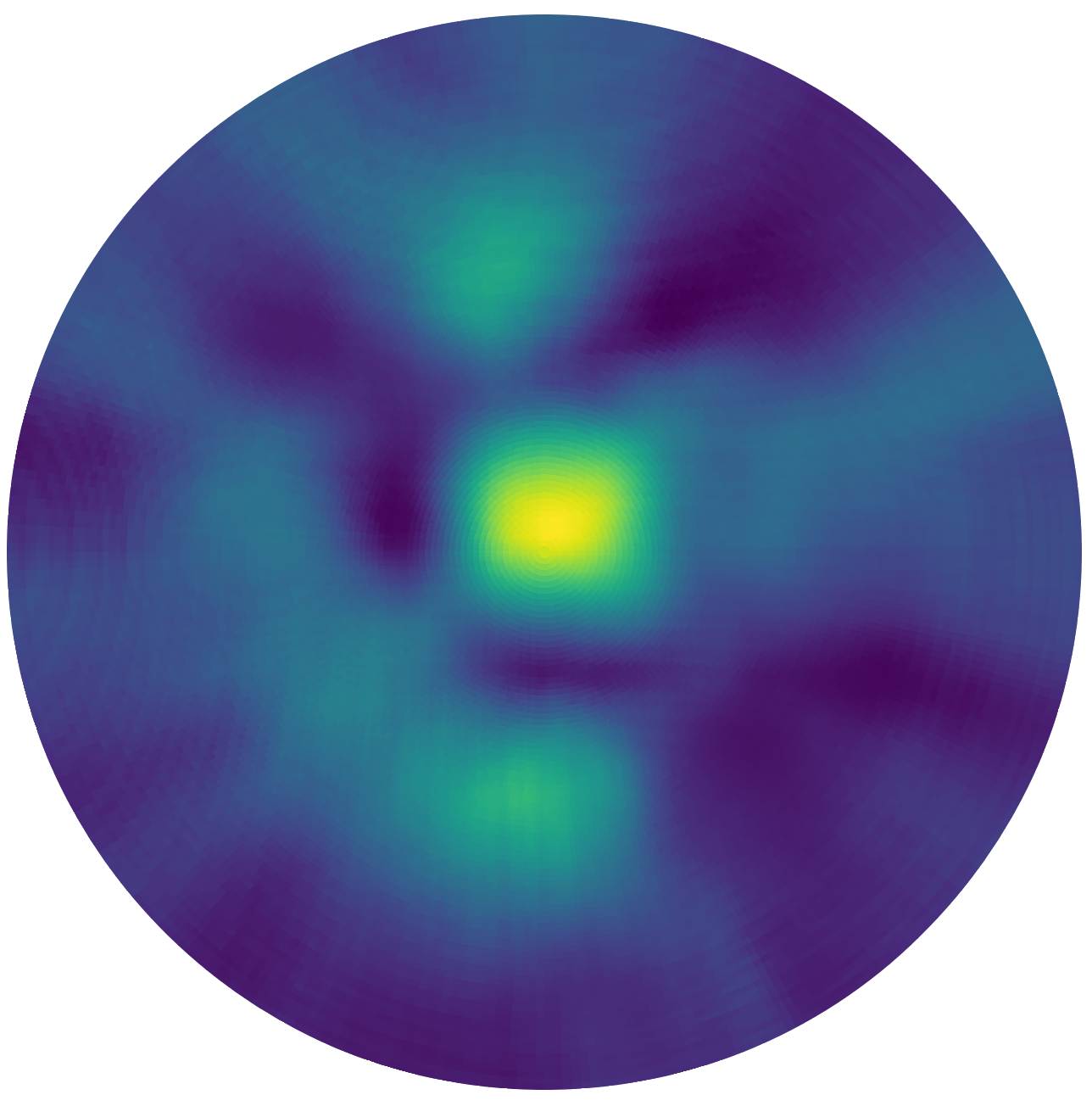}{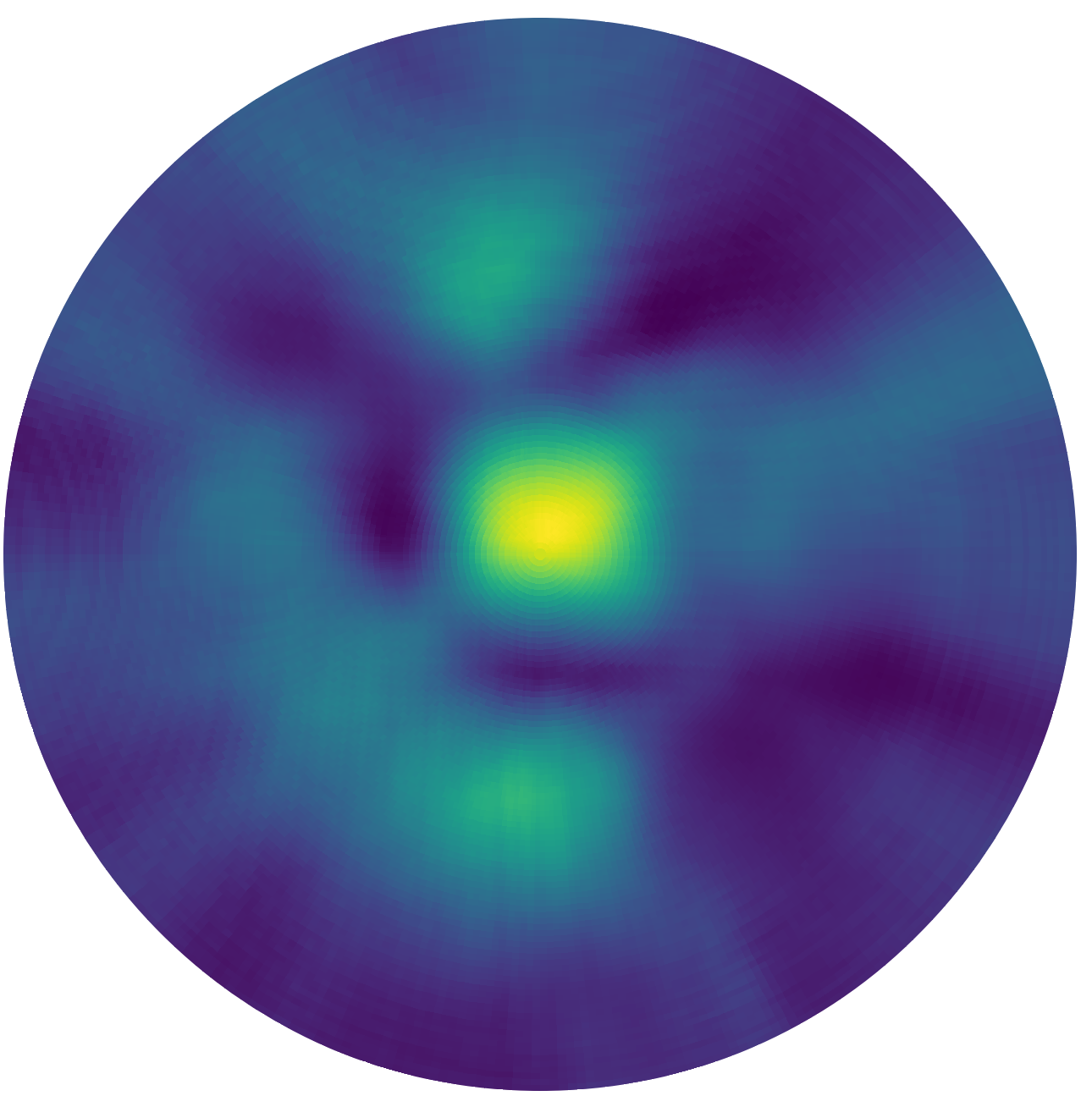} &
    \colstack{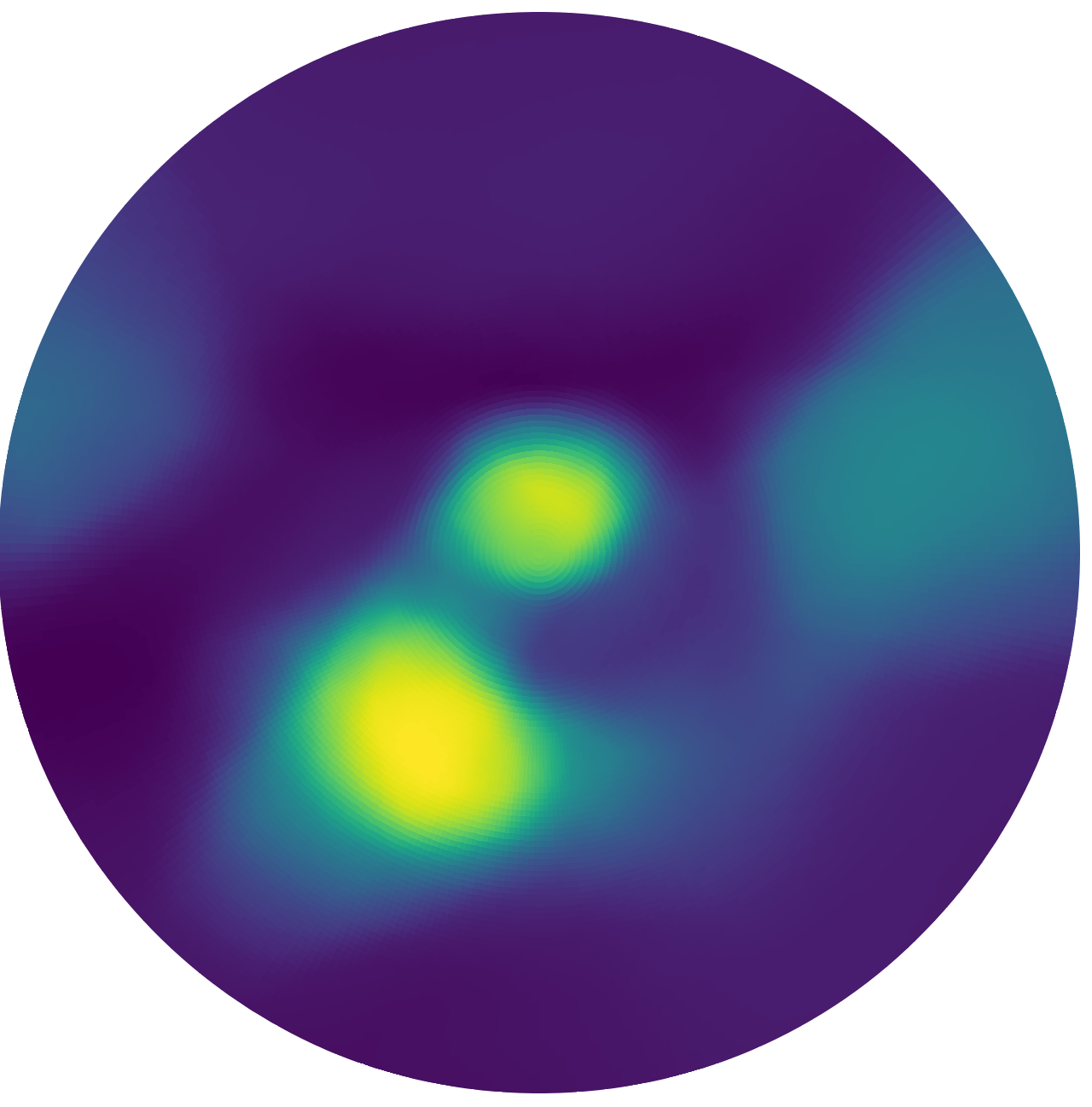}{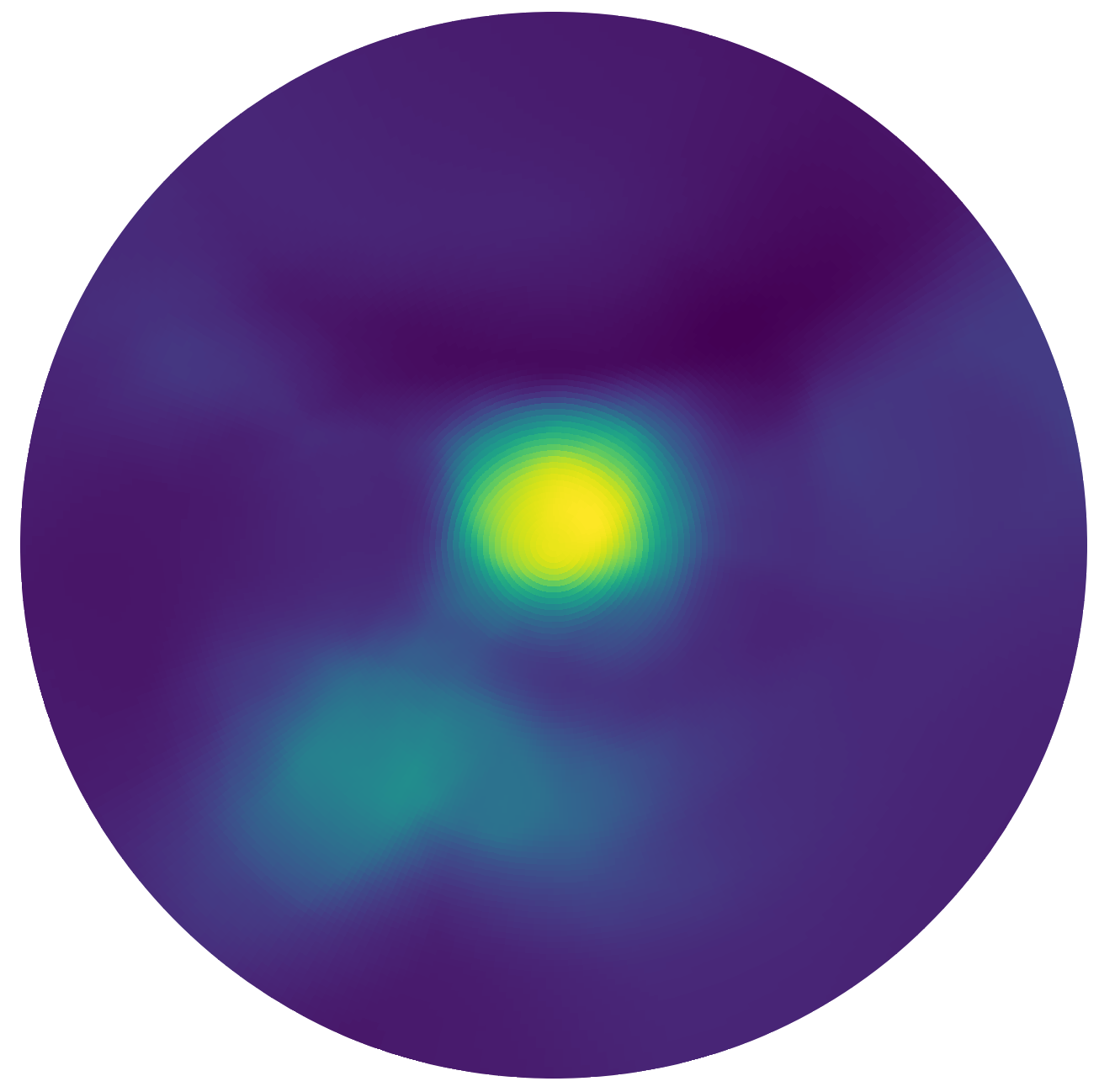}{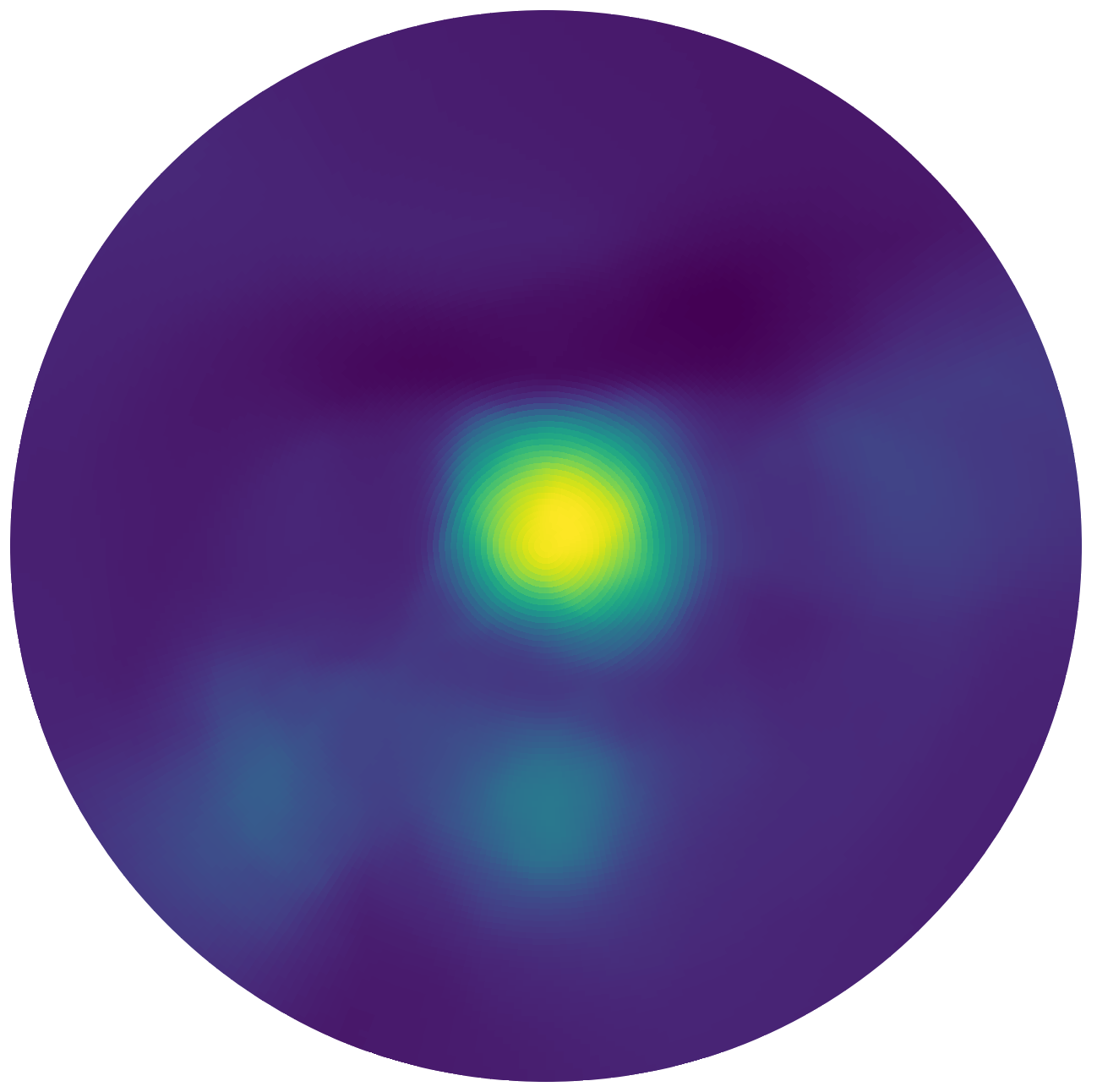} &
    \colstack{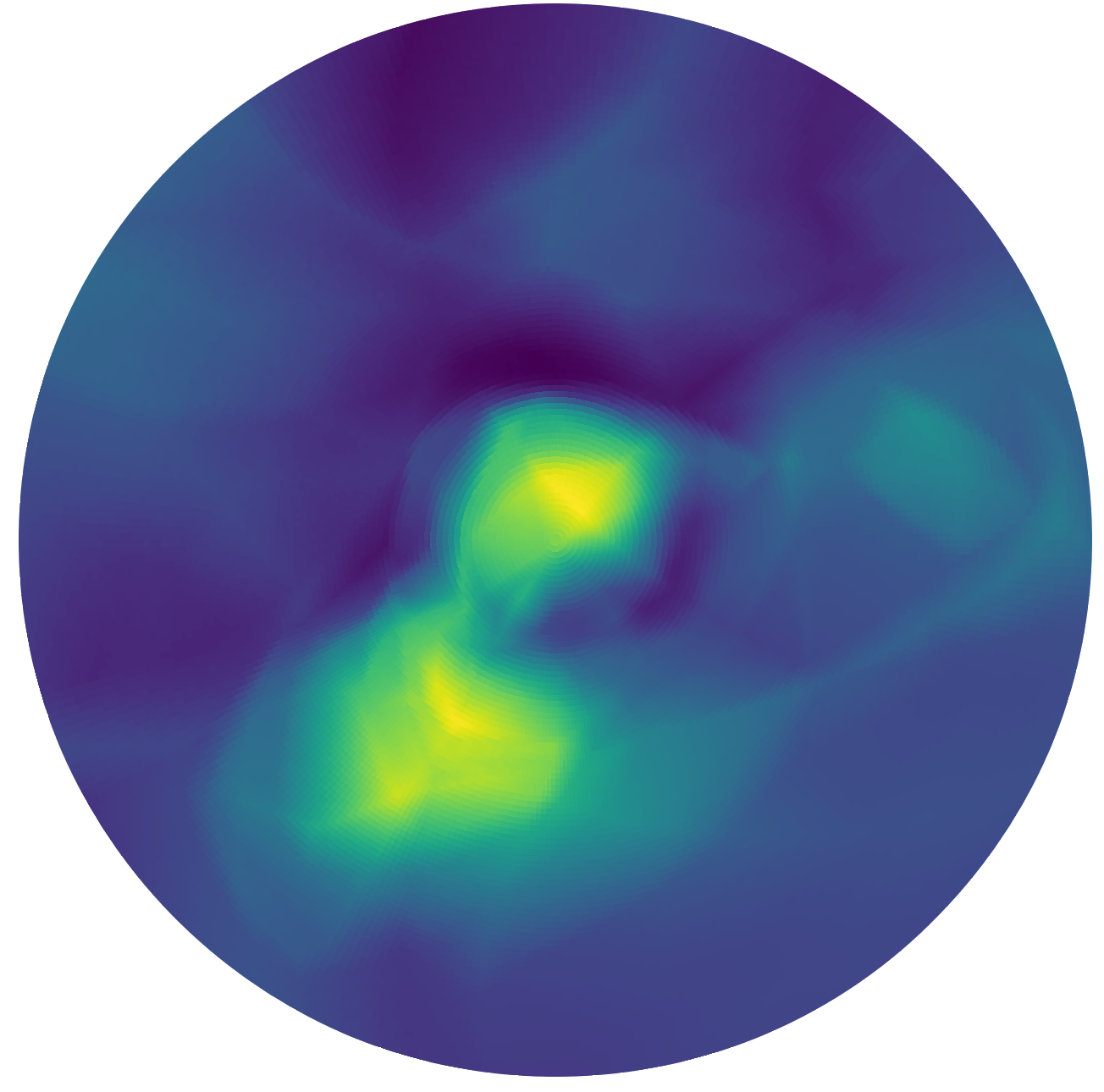}{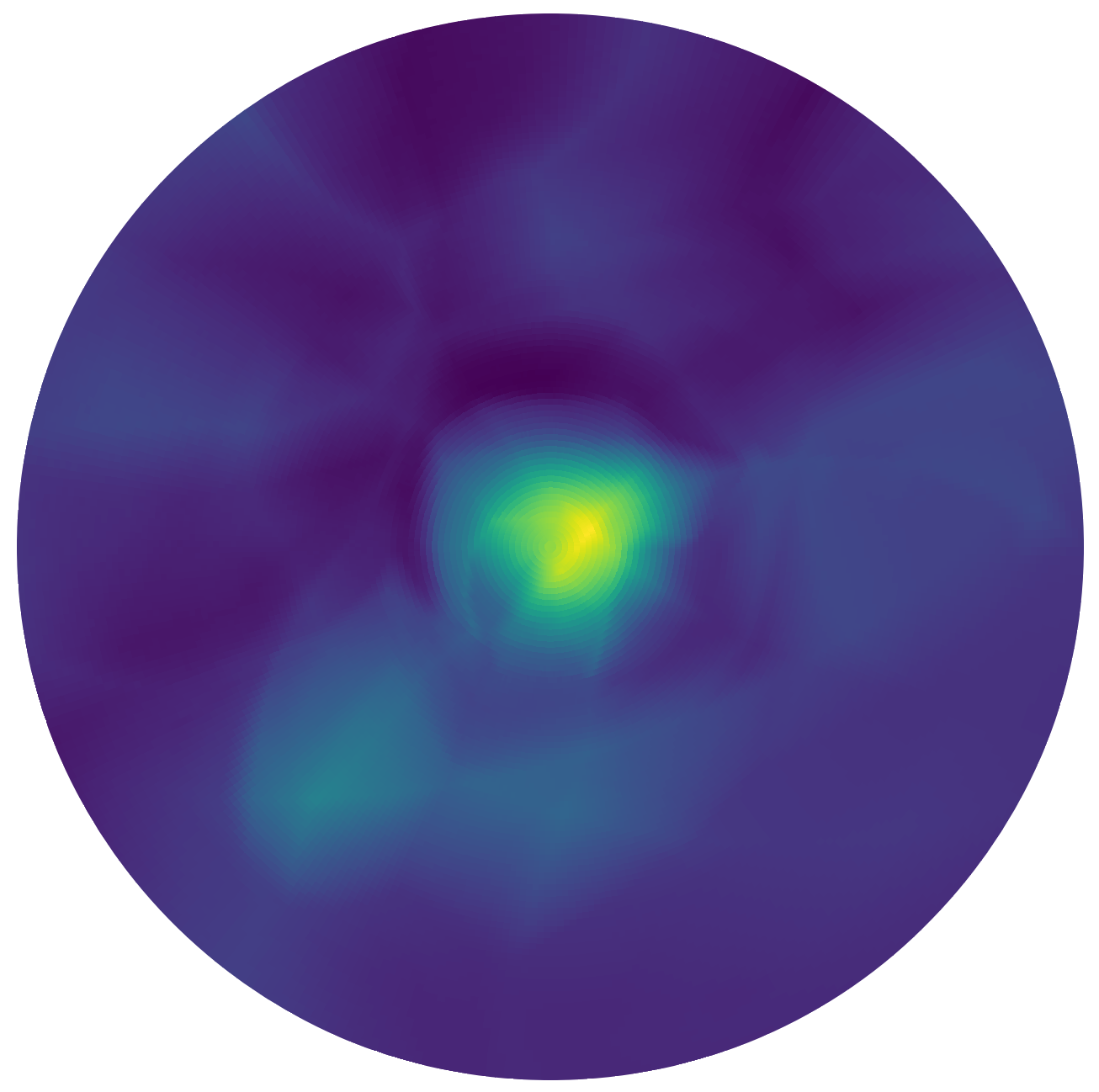}{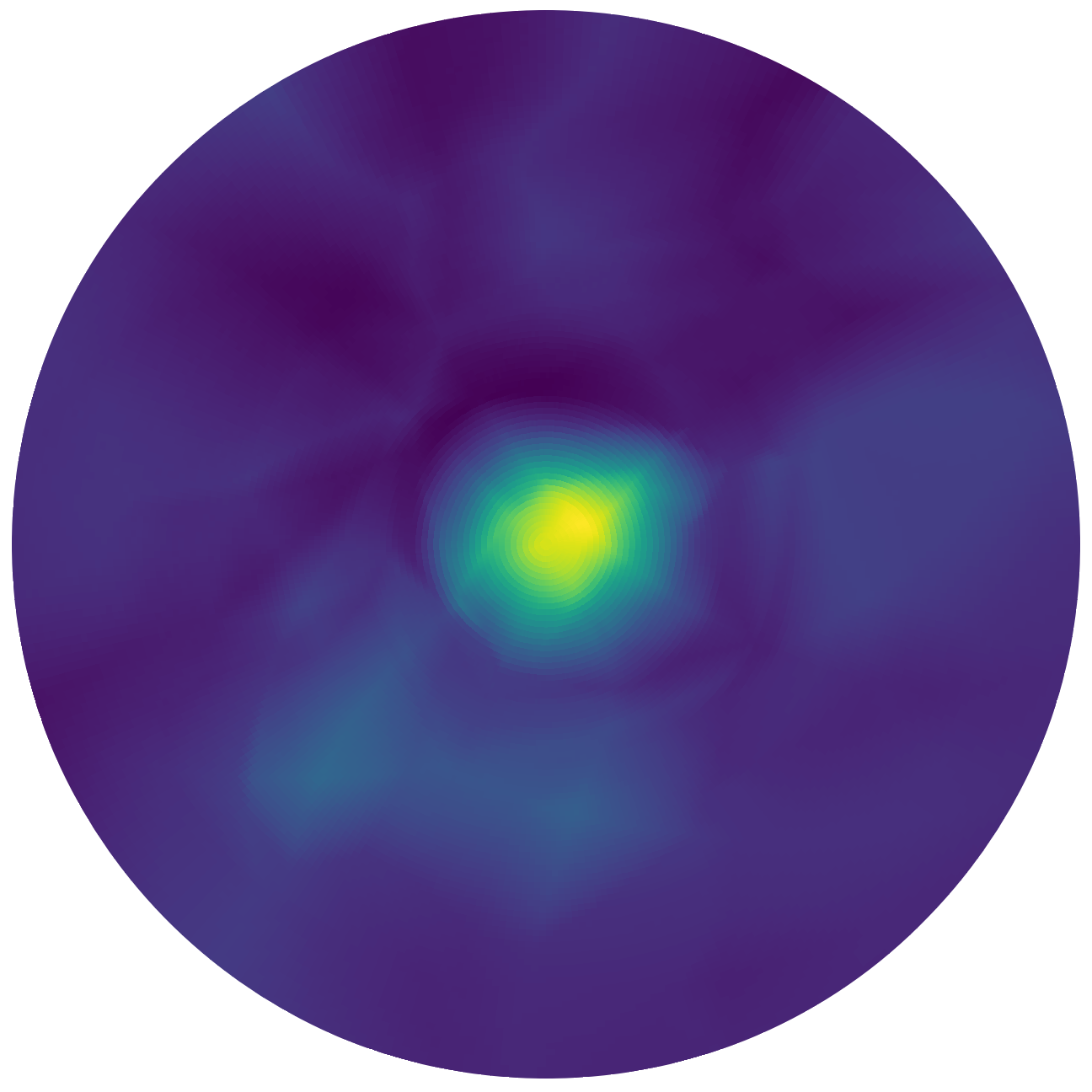} &
    \colstack{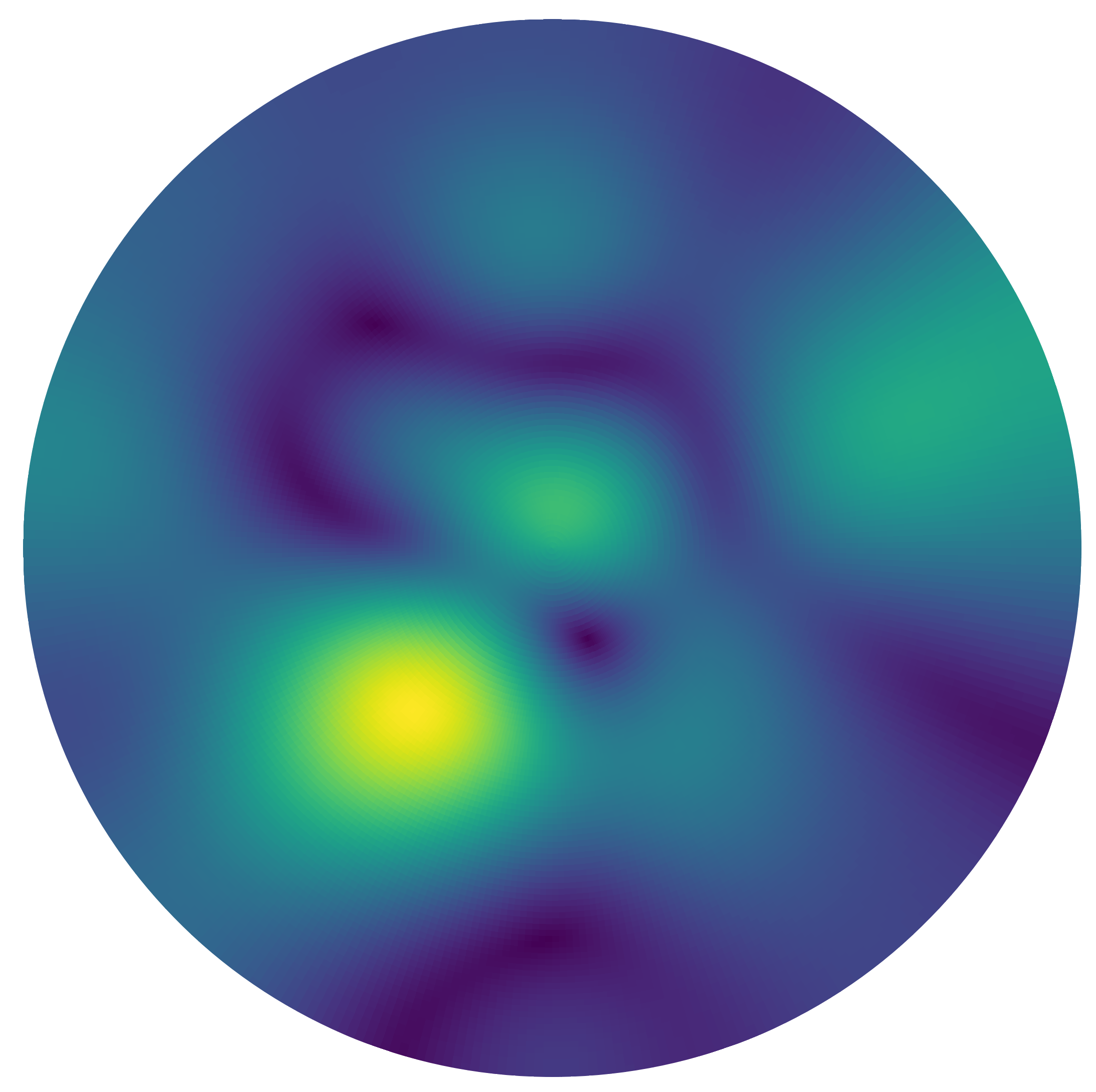}{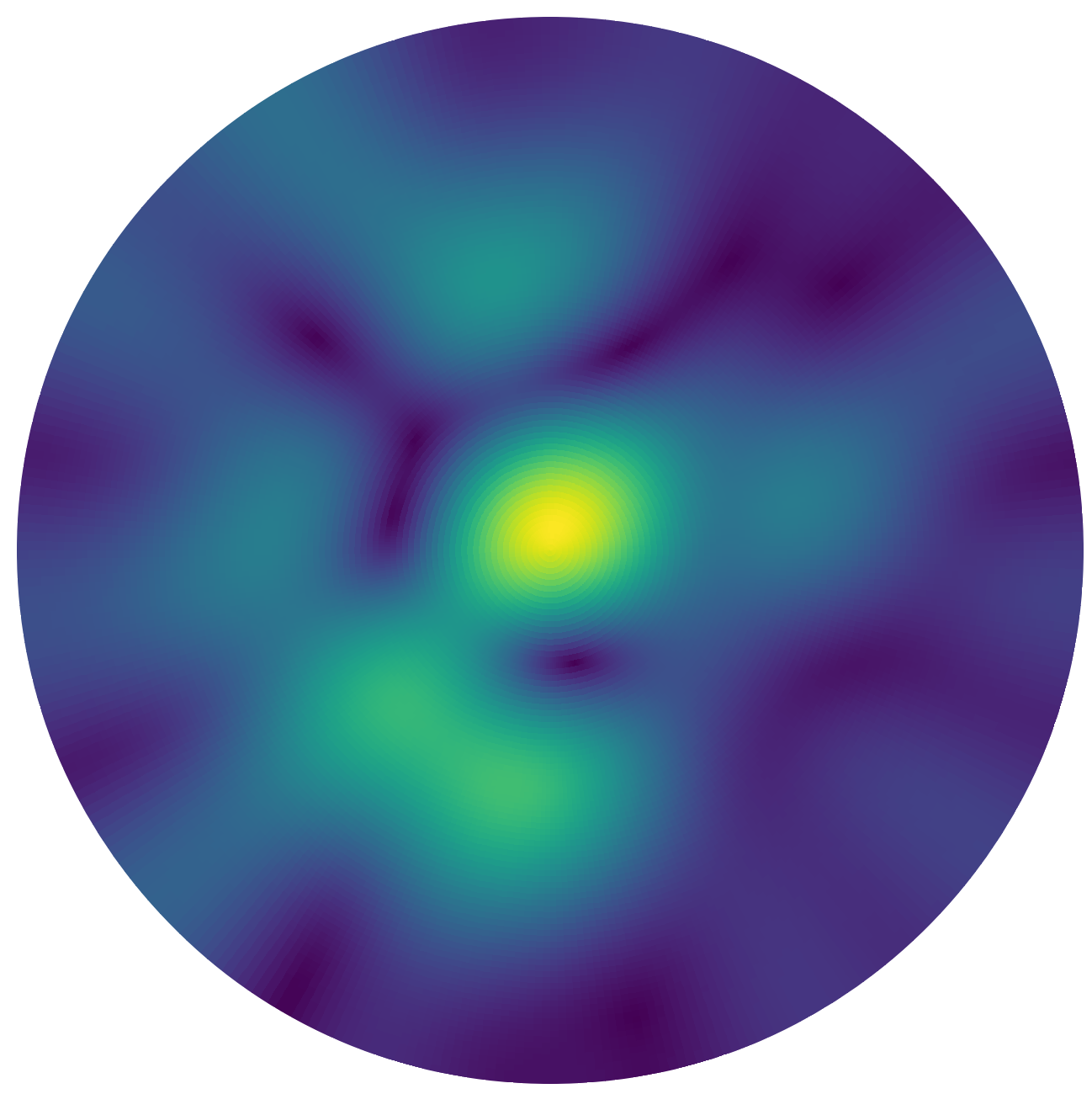}{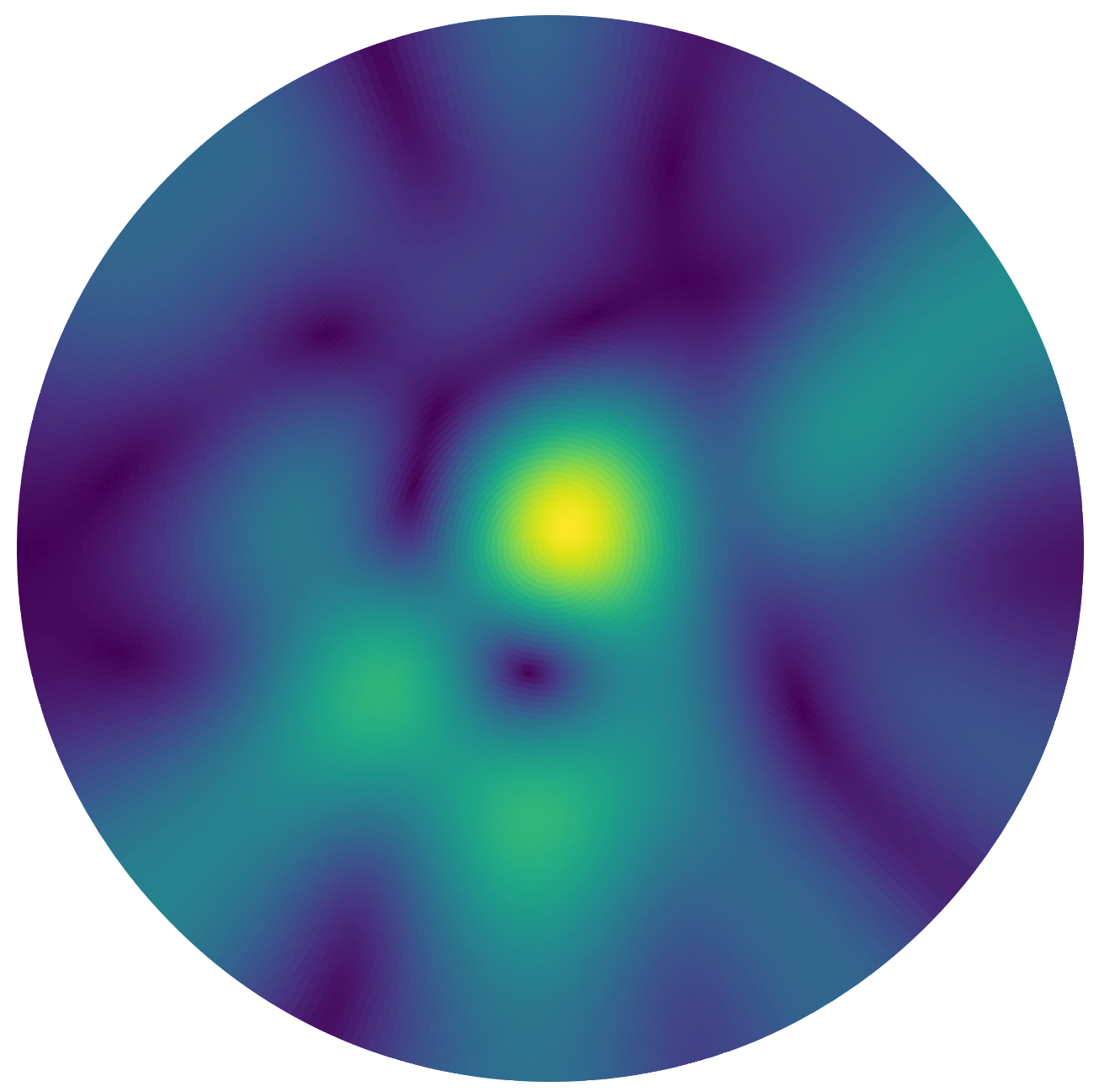}
  \end{tabular}

  \vspace{3pt}

  \begin{tabular}{@{}c@{\hspace{\colgap}}c@{\hspace{\colgap}}c@{\hspace{\colgap}}c@{\hspace{\colgap}}c@{\hspace{\colgap}}c@{\hspace{\colgap}}c@{\hspace{\colgap}}c@{}}
    \makebox[\colw][c]{(a) Ours} &
    \makebox[\colw][c]{(b) WRF-GS} &
    \makebox[\colw][c]{(c) DCGAN} &
    \makebox[\colw][c]{(d) NeRF$^2$} &
    \makebox[\colw][c]{(e) NeRF-APT} &
    \makebox[\colw][c]{(f) MLP} &
    \makebox[\colw][c]{(g) FIRE} &
    \makebox[\colw][c]{(h) Label}
  \end{tabular}

  \caption{2D spatial spectrum visualizations.}
  \label{fig:2dvisualization}
\end{figure*}

\section{Experimental Evaluation and Analysis}
\label{sec:eva_ana}

\begin{figure}[htbp]
\centering
\includegraphics[width=\linewidth]{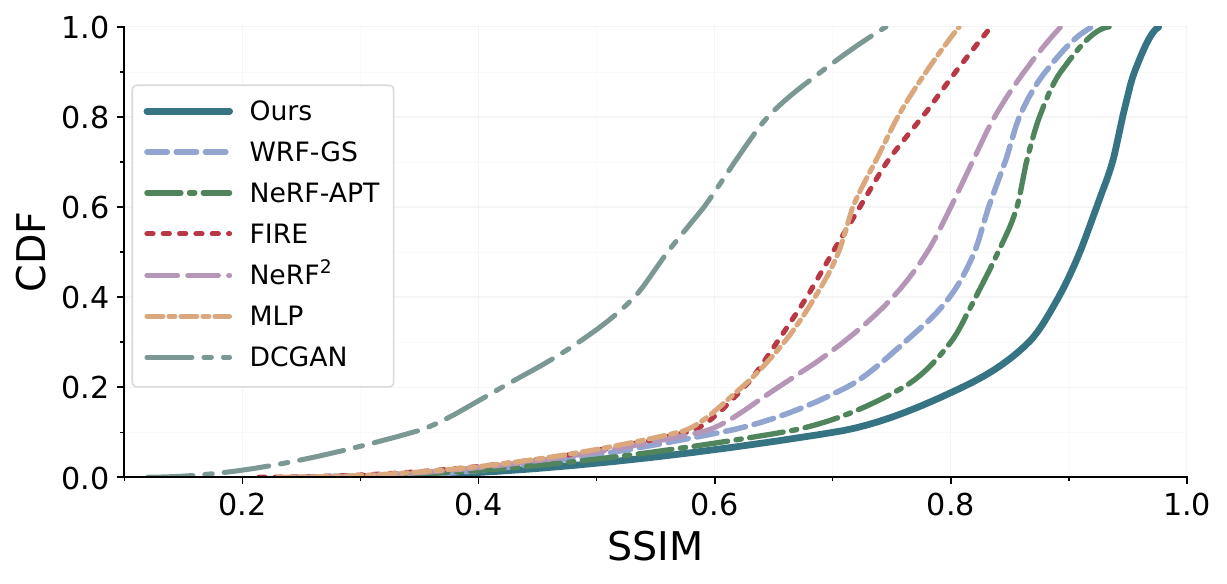}
\vspace{-2em}
\caption{CDF-SSIM comparison.}
\label{fig:figure_cdf}
\end{figure}

\begin{figure}[!htbp]
    \centering
    \begin{subfigure}[t]{0.33\linewidth}
        \centering
        \includegraphics[width=\linewidth]{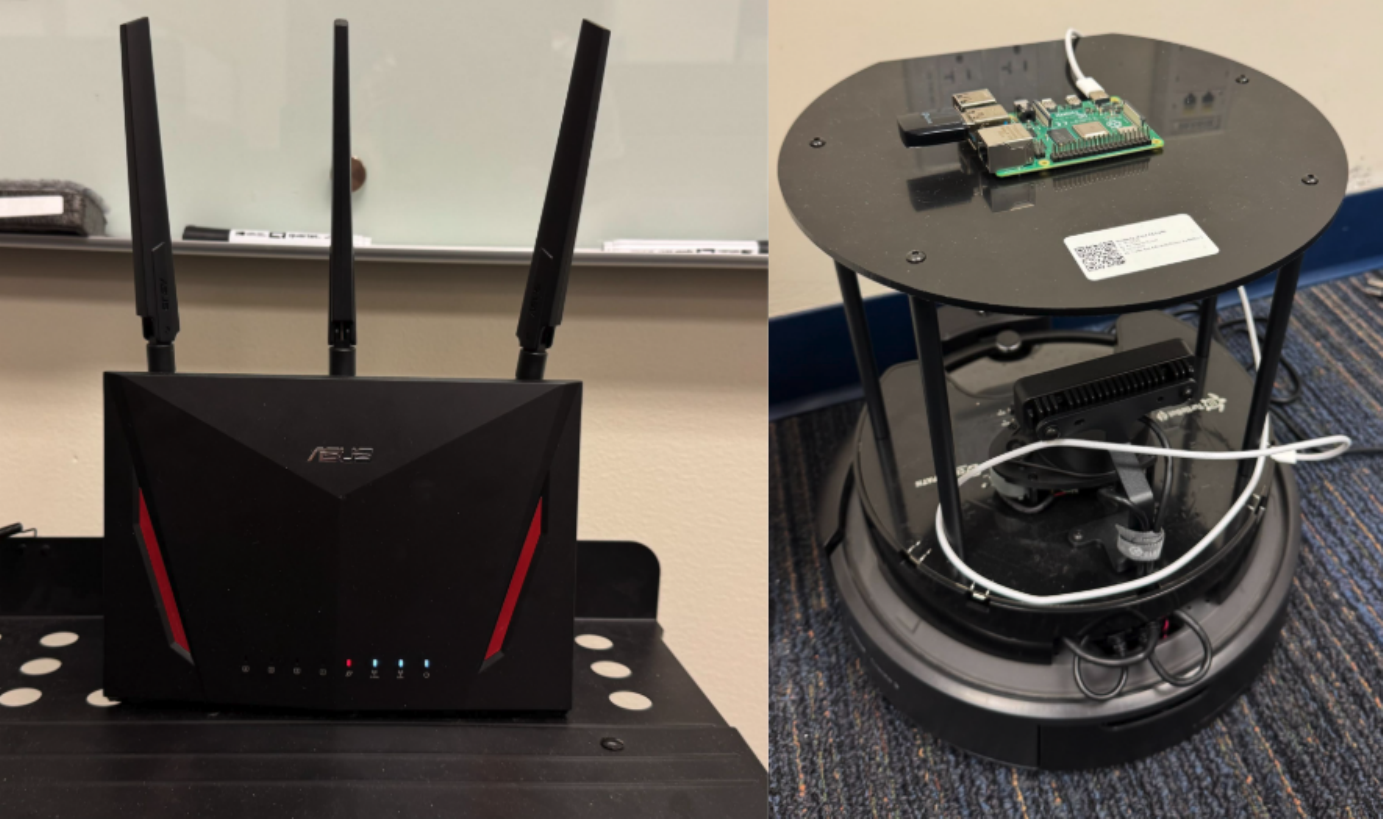}
        \caption{Platforms.}
        \label{fig:devices}
    \end{subfigure}
    \hfill
    \begin{subfigure}[t]{0.66\linewidth}
        \centering
        \includegraphics[width=\linewidth]{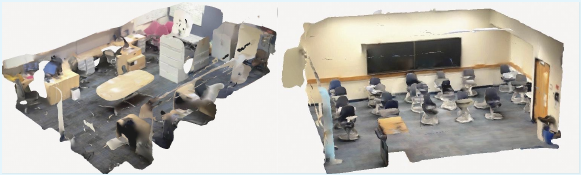}
        \caption{Room 1 and Room 2 overview.}
        \label{fig:room_overview}
    \end{subfigure}

    \caption{Overview of the platforms and room configurations.}
    \label{fig:overview_all}
    \vspace{-0.8em}
\end{figure}

\subsection{Datasets}
We built a semi-automated platform to map indoor RSSI at 2.4 GHz and 5 GHz over an approximately $35 \mathrm{m}^2$ site during an 18-day campaign. As shown in Fig.~\ref{fig:devices}, a TurtleBot 4 carrying a Raspberry Pi 4 measurement unit was equipped with an RPLIDAR A1M8 and the Nav2+AMCL stack, providing real-time SLAM, an occupancy grid, and precise $(x,y,z)$ poses for spatially referenced measurements. The Pi device hosted a dual-band TP-Link Archer T3U AC1300 adapter, while two ASUS RT-AC86U routers provided infrastructure: one acted as the transmitter, fixed at 1 m height in dual-band mode with maximum power, and the other maintained system communication on non-overlapping channels. Data collection combined joystick navigation to predefined waypoints with automated acquisition: at each location we logged the pose and recorded five RSSI samples per band at 1 s intervals, accepting a measurement only if all five reads were successfully obtained. We consider two measured rooms and their reference maps, both of which contain structural columns that create clear occlusions useful for analyzing ray–object interactions. Fig.~\ref{fig:room_overview} shows the overview of two different rooms.

In addition, we evaluate on the public BLE and the RFID spectrum datasets from the NeRF$^2$~\cite{10.1145/3570361.3592527} repositories. In the BLE dataset, each sample is a 50-dimensional RSSI vector from the gateways plus the tag position. We use 4.2k samples for training and 1.8k for testing. In the RFID spectrum dataset, array snapshots are converted to a front-hemisphere spatial spectrum on a $360\times90$ azimuth–elevation grid.

\subsection{Baselines}
We benchmark against several primary baselines and one ray-tracing variant. First, we use MATLAB’s Ray Tracing toolbox to obtain the corresponding results~\cite{7152831}. Specifically, the toolbox requires a 3D scene model as input and predicts the RF signals at the Rx locations given the Tx positions. Second, an MLP with four hidden layers followed by a linear output head. Third, we use FIRE~\cite{10.1145/3447993.3483275} with a three-layer encoder and a symmetric four-layer decoder to evaluate on our proposed datasets. Fourth, a DCGAN network~\cite{radford2015unsupervised} in which both generator and discriminator have four layers. Next, the standard NeRF$^2$~\cite{10.1145/3570361.3592527} configuration implemented as in the original release. For the sixth baseline evaluation, we use the NeRF-APT~\cite{shen2025nerfaptnewnerfframework} which is a variant of NeRF$^2$. Furthermore, we use the state-of-the-art model WRF-GS~\cite{11044513} as our final baseline.

\begin{table*}[t]
\centering
\caption{Performance comparison of GAI-GS against baseline methods across different room configurations and BLE-RSSI/RFID spectrum datasets. Lower MAE values and higher SSIM values indicate better performance.}
\resizebox{0.75\linewidth}{!}{
\begin{tabular}{c|cc|cc|cc}
\toprule
\multirow{3}{*}{\textbf{Method}} & \multicolumn{2}{c|}{\textbf{Room 1 (Our RSSI Dataset)}} & \multicolumn{2}{c|}{\textbf{Room 2 (Our RSSI Dataset)}} & \multicolumn{2}{c}{\textbf{Other Dataset}} \\ \cline{2-7} 
                                 & \multicolumn{2}{c|}{\textbf{MAE (dB) $\downarrow$}}                  & \multicolumn{2}{c|}{\textbf{MAE (dB) $\downarrow$}}                  & \textbf{MAE (dB) $\downarrow$}    & \textbf{SSIM $\uparrow$}       \\ \cline{2-7} 
                                 & \textbf{2.4 GHz}           & \textbf{5.0 GHz}           & \textbf{2.4 GHz}           & \textbf{5.0 GHz}           & \textbf{BLE}    & \textbf{Spectrum}   \\ \midrule
Ray Tracing~\cite{7152831}                      & 25.52                      & 20.66                      & 25.56                      & 20.78                      & --              & 0.33                  \\
MLP                              & 7.3                        & 9.3                        & 8.2                        & 9.9                        & 8.0             & 0.71                \\
FIRE~\cite{10.1145/3447993.3483275}
                                 & 5.8                        & 5.5                        & 4.5                        & 2.7                        & 6.4             & 0.73                \\
DCGAN~\cite{radford2015unsupervised}
                                 & 4.0                        & 3.4                        & 4.2                        & 3.0                        & 4.6             & 0.56                \\
NeRF$^2$~\cite{10.1145/3570361.3592527}
                                 & 3.6                        & 2.9                        & 3.9                        & 2.0                        & 3.1             & 0.78                \\
NeRF-APT~\cite{shen2025nerfaptnewnerfframework}
                                 & 3.3                        & 2.7                        & 3.6                        & 2.0                        & 3.1             & 0.84                \\
WRF-GS~\cite{11044513}           & 3.1                        & 2.4                        & 3.1                        & 1.9                        & 2.8             & 0.82                \\ \midrule

\textbf{GAI-GS}                  & \textbf{2.9}               & \textbf{1.6}               & \textbf{2.7}               & \textbf{1.8}               & \textbf{2.3}    & \textbf{0.91}       \\ 
\bottomrule
\end{tabular}}
\label{tab:performance_comparison}
\end{table*}

\subsection{Results}
We use a combination of the mean absolute error (MAE) and the structural similarity (SSIM) loss between predicted results and labels, with an $L_2$ penalty on $d_{attn}$ for the spectrum dataset:
\begin{multline}
    L = \frac{1}{M} \sum_{i=1}^{M} \big( \beta L_{\text{MAE}}(I_{gt}, I_{pred})\\
    + (1 - \beta) L_{\text{SSIM}}(I_{gt}, I_{pred}) \big)
    + \alpha L_{\text{2}}(d_{attn}),
\end{multline}
where $I_{gt}, I_{pred}$ are the ground truth and synthesized spatial spectrum, while $M$ is the number of measurements and $\beta$ and $\alpha$ are weight parameters. As for the BLE datasets, measurements corresponding to receiver locations at excessive distances from the transmitter are treated as invalid and excluded from our analysis, as these instances yield degenerate RSSI recordings of $-100$ dBm. The remaining valid measurements are subsequently partitioned into a training/evaluation split. The reported metric, expressed in dB, represents the median MAE aggregated across all receivers in the test set. 

We compare training, inference, and rendering times against prior 3DGS-based wireless models on a single NVIDIA A100 GPU using a subset of the Spectrum dataset, shown in Table~\ref{tab:efficiency}. Despite requiring additional computation during training and inference, our method provides better reconstruction quality while also outperforming prior methods in rendering speed.

\begin{figure}[htbp]
\centering
\includegraphics[width=\linewidth]{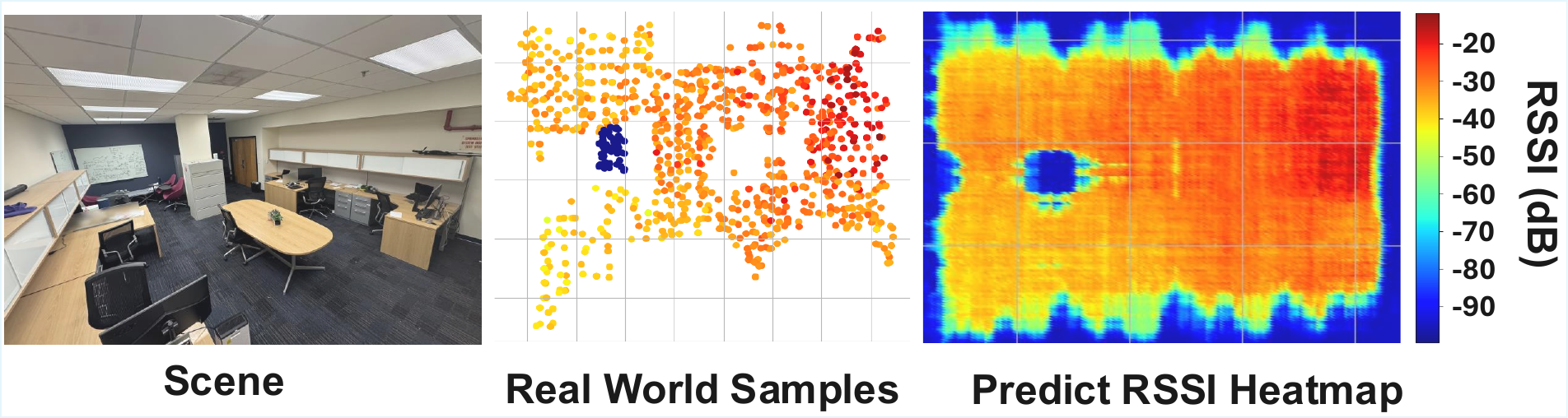}
\caption{Visualizations of RSSI prediction.}
\label{figure_vis}
\end{figure}

Table~\ref{tab:performance_comparison} summarizes the quantitative results across all datasets and room configurations. GAI-GS consistently achieves the lowest MAE and the highest SSIM among all compared methods. On our RSSI dataset in Room~1, GAI-GS reduces the MAE to 2.9~dB at 2.4~GHz and 1.6~dB at 5.0~GHz, improving over the strongest baseline WRF-GS by 1.2~dB and 0.6~dB, respectively. In Room~2, GAI-GS achieves 2.7~dB and 1.8~dB at the two frequency bands, surpassing WRF-GS by 0.4~dB and 0.1~dB. Among the baselines, classical methods such as Ray Tracing and MLP show limited capacity with MAE values exceeding 7.0~dB. Learning-based approaches progressively improve, with FIRE reaching 2.7--5.8~dB, DCGAN 3.0--4.2~dB, and neural radiance field methods narrowing the gap to 2.0--3.9~dB. GAI-GS outperforms all of them by a clear margin, demonstrating the effectiveness of our GA-informed interaction modeling. On the external BLE-RSSI dataset, GAI-GS attains an MAE of 2.3~dB, outperforming WRF-GS at 2.8~dB and NeRF-based methods at 3.1~dB. For spectrum reconstruction, GAI-GS reaches an SSIM of 0.91, a substantial improvement over NeRF-APT at 0.84 and WRF-GS at 0.82. These results confirm that GAI-GS achieves superior spatial accuracy and perceptual fidelity across all evaluated settings.

Fig.~\ref{fig:2dvisualization} presents the 2D spatial spectrum visualizations comparing our method with other baselines. MLP and FIRE exhibit blurred and distorted spectral patterns, indicating a loss of spatial coherence. WRF-GS achieves relatively better results but fails to preserve high-frequency structural details.
In contrast, our method produces the most accurate and physically consistent spectra, closely matching the ground-truth label. The reconstructed spectra exhibit sharp high-energy focal regions and fine-grained propagation textures that reflect realistic wave behaviors. This demonstrates that our GAI-GS approach effectively encodes spatial correlations and physical consistency, leading to more faithful recovery of the underlying EM field structure.

Fig.~\ref{fig:figure_cdf} shows the cumulative distribution function (CDF) of the SSIM for different methods. A curve shifted toward the upper right indicates better overall perceptual quality and higher structural consistency. Also, Fig.~\ref{figure_vis} illustrates the visualization of predicted RSSI maps with the samples extracted from the scene.

Overall, these results show that our GAI-GS framework enhances spatial representation fidelity and interaction-aware modeling, leading to more accurate EM propagation modeling.

\begin{table}[h]
    \centering
    \footnotesize
    \resizebox{\linewidth}{!}{
        \begin{tabular}{lccc}
            \toprule
            Method & Training (mins) & Inference (ms) & Render (ms) \\
            \midrule
            WRF-GS~\cite{11044513}  & 312.38 & 434.21 & 39.29 \\
            WRF-GS+~\cite{wen2024neural} & \textbf{101.06} & \textbf{4.78} & 1.43 \\
            \textbf{Ours} & 203.00 & 17.37 & \textbf{0.91} \\
            \bottomrule
        \end{tabular}
    }
    \caption{Comparisons of training, inference, and rendering time.}
    \vspace{-2em}
    \label{tab:efficiency}
\end{table}
\section{Conclusion}
\label{sec:conclusion}

In this paper, we propose GAI-GS, a geometric algebra–informed 3D Gaussian splatting framework that unifies geometric algebra and wireless ray-transmission theory for wireless modeling. By coupling a 3D-GS scene representation with physically grounded propagation mechanisms, GAI-GS implicitly captures ray–object interactions and enables more accurate reconstruction of the wireless field. Extensive experiments on both public and in-room datasets demonstrate its effectiveness and consistent gains over baselines. We believe that GAI-GS provides a more principled basis for wireless channel modeling and offers a flexible foundation for scaling 3D-GS–based methods to increasingly complex wireless communication environments. 

\section*{Acknowledgments}
This work is supported in part by the U.S. NSF under grants (CNS-2415209, CNS-2317190, IIS-2306791, and CNS-2319343).

{
    \small
    \bibliographystyle{ieeenat_fullname}
    \bibliography{main}

\begin{thebibliography}{45}
\providecommand{\natexlab}[1]{#1}
\providecommand{\url}[1]{\texttt{#1}}
\expandafter\ifx\csname urlstyle\endcsname\relax
  \providecommand{\doi}[1]{doi: #1}\else
  \providecommand{\doi}{doi: \begingroup \urlstyle{rm}\Url}\fi

\bibitem[Aldossari and Chen(2019)]{aldossari2019machine}
Saud~Mobark Aldossari and Kwang-Cheng Chen.
\newblock Machine learning for wireless communication channel modeling: An overview.
\newblock \emph{Wireless Personal Communications}, 106\penalty0 (1):\penalty0 41--70, 2019.

\bibitem[Almers et~al.(2007)Almers, Bonek, Burr, Czink, Debbah, Degli-Esposti, Hofstetter, Ky{\"o}sti, Laurenson, Matz, et~al.]{almers2007survey}
Peter Almers, Ernst Bonek, Alister Burr, Nicolai Czink, M{\'e}rouane Debbah, Vittorio Degli-Esposti, Helmut Hofstetter, Pekka Ky{\"o}sti, David Laurenson, Gerald Matz, et~al.
\newblock Survey of channel and radio propagation models for wireless {MIMO} systems.
\newblock \emph{EURASIP Journal on Wireless Communications and Networking}, 2007\penalty0 (1):\penalty0 019070, 2007.

\bibitem[Arvinte and Tamir(2022)]{9771907}
Marius Arvinte and Jonathan~I. Tamir.
\newblock Score-based generative models for robust channel estimation.
\newblock In \emph{2022 IEEE Wireless Communications and Networking Conference (WCNC)}, pages 453--458, 2022.

\bibitem[Bakshi et~al.(2019)Bakshi, Mao, Srinivasan, and Parthasarathy]{10.1145/3300061.3345438}
Arjun Bakshi, Yifan Mao, Kannan Srinivasan, and Srinivasan Parthasarathy.
\newblock Fast and efficient cross band channel prediction using machine learning.
\newblock In \emph{The 25th Annual International Conference on Mobile Computing and Networking}, pages 1--16, 2019.

\bibitem[Brehmer et~al.(2023)Brehmer, De~Haan, Behrends, and Cohen]{brehmer2023geometric}
Johann Brehmer, Pim De~Haan, S{\"o}nke Behrends, and Taco~S Cohen.
\newblock Geometric algebra transformer.
\newblock \emph{Advances in Neural Information Processing Systems}, 36:\penalty0 35472--35496, 2023.

\bibitem[Chen et~al.(2025)Chen, Feng, Qian, and Zhang]{chen2025radio}
Xingyu Chen, Zihao Feng, Kun Qian, and Xinyu Zhang.
\newblock Radio frequency ray tracing with neural object representation for enhanced {RF} modeling.
\newblock In \emph{Proceedings of the Computer Vision and Pattern Recognition Conference}, pages 21339--21348, 2025.

\bibitem[Chen et~al.(2024)Chen, Xu, Zheng, Zhuang, Pollefeys, Geiger, Cham, and Cai]{chen2024mvsplat}
Yuedong Chen, Haofei Xu, Chuanxia Zheng, Bohan Zhuang, Marc Pollefeys, Andreas Geiger, Tat-Jen Cham, and Jianfei Cai.
\newblock Mvsplat: Efficient {3D} gaussian splatting from sparse multi-view images.
\newblock In \emph{European Conference on Computer Vision}, pages 370--386. Springer, 2024.

\bibitem[Devlin et~al.(2019)Devlin, Chang, Lee, and Toutanova]{47751}
Jacob Devlin, Ming-Wei Chang, Kenton Lee, and Kristina Toutanova.
\newblock Bert: Pre-training of deep bidirectional transformers for language understanding.
\newblock In \emph{Proceedings of the 2019 conference of the North American chapter of the association for computational linguistics: human language technologies, volume 1 (long and short papers)}, pages 4171--4186, 2019.

\bibitem[Dorst and De~Keninck(2022)]{dorst2022guided}
Leo Dorst and Steven De~Keninck.
\newblock A guided tour to the plane-based geometric algebra pga.
\newblock \emph{URL https://bivector. net/PGA4CS. html}, 2022.

\bibitem[He et~al.(2019)He, Ai, Guan, Wang, Zhong, and Kürner]{8438326}
Danping He, Bo Ai, Ke Guan, Longhe Wang, Zhangdui Zhong, and Thomas Kürner.
\newblock The design and applications of high-performance ray-tracing simulation platform for {5G} and beyond wireless communications: A tutorial.
\newblock \emph{IEEE Communications Surveys \& Tutorials}, 21\penalty0 (1):\penalty0 10--27, 2019.

\bibitem[He et~al.(2018)He, Ai, Molisch, Stuber, Li, Zhong, and Yu]{he2018clustering}
Ruisi He, Bo Ai, Andreas~F Molisch, Gordon~L Stuber, Qingyong Li, Zhangdui Zhong, and Jian Yu.
\newblock Clustering enabled wireless channel modeling using big data algorithms.
\newblock \emph{IEEE Communications Magazine}, 56\penalty0 (5):\penalty0 177--183, 2018.

\bibitem[Huang et~al.(2018)Huang, Wang, Bai, Sun, Yang, Li, Tirkkonen, and Zhou]{huang2018big}
Jie Huang, Cheng-Xiang Wang, Lu Bai, Jian Sun, Yang Yang, Jie Li, Olav Tirkkonen, and Ming-Tuo Zhou.
\newblock A big data enabled channel model for {5G} wireless communication systems.
\newblock \emph{IEEE Transactions on Big Data}, 6\penalty0 (2):\penalty0 211--222, 2018.

\bibitem[Imoize et~al.(2021)Imoize, Ibhaze, Atayero, and Kavitha]{imoize2021standard}
Agbotiname~Lucky Imoize, Augustus~Ehiremen Ibhaze, Aderemi~A Atayero, and KVN Kavitha.
\newblock Standard propagation channel models for {MIMO} communication systems.
\newblock \emph{Wireless Communications and Mobile Computing}, 2021\penalty0 (1):\penalty0 8838792, 2021.

\bibitem[Jia et~al.(2024)Jia, Chen, Wei, Sun, and Pi]{jia2025neuralreflectancefieldsradiofrequency}
Haifeng Jia, Xinyi Chen, Yichen Wei, Yifei Sun, and Yibo Pi.
\newblock Neural reflectance fields for radio-frequency ray tracing.
\newblock In \emph{GLOBECOM 2024-2024 IEEE Global Communications Conference}, pages 4226--4231. IEEE, 2024.

\bibitem[Jiang et~al.(2016)Jiang, Zhang, Ren, Han, Chen, and Hanzo]{7792374}
Chunxiao Jiang, Haijun Zhang, Yong Ren, Zhu Han, Kwang-Cheng Chen, and Lajos Hanzo.
\newblock Machine learning paradigms for next-generation wireless networks.
\newblock \emph{IEEE Wireless Communications}, 24\penalty0 (2):\penalty0 98--105, 2016.

\bibitem[Jiang et~al.(2025)Jiang, Sivaram, Peng, and Ramamoorthi]{jiang2025geometry}
Kaiwen Jiang, Venkataram Sivaram, Cheng Peng, and Ravi Ramamoorthi.
\newblock Geometry field splatting with {Gaussian} surfels.
\newblock In \emph{Proceedings of the Computer Vision and Pattern Recognition Conference}, pages 5752--5762, 2025.

\bibitem[Kerbl et~al.(2023)Kerbl, Kopanas, Leimk{\"u}hler, and Drettakis]{kerbl20233d}
Bernhard Kerbl, Georgios Kopanas, Thomas Leimk{\"u}hler, and George Drettakis.
\newblock {3D} gaussian splatting for real-time radiance field rendering.
\newblock \emph{ACM Trans. Graph.}, 42\penalty0 (4):\penalty0 139--1, 2023.

\bibitem[Li et~al.(2022)Li, Li, Xiong, and Hoi]{pmlr-v162-li22n}
Junnan Li, Dongxu Li, Caiming Xiong, and Steven Hoi.
\newblock Blip: Bootstrapping language-image pre-training for unified vision-language understanding and generation.
\newblock In \emph{International conference on machine learning}, pages 12888--12900. PMLR, 2022.

\bibitem[Li et~al.(2023)Li, Li, Savarese, and Hoi]{pmlr-v202-li23q}
Junnan Li, Dongxu Li, Silvio Savarese, and Steven Hoi.
\newblock Blip-2: Bootstrapping language-image pre-training with frozen image encoders and large language models.
\newblock In \emph{International conference on machine learning}, pages 19730--19742. PMLR, 2023.

\bibitem[Liu et~al.(2021)Liu, Singh, Xu, and Vasisht]{10.1145/3447993.3483275}
Zikun Liu, Gagandeep Singh, Chenren Xu, and Deepak Vasisht.
\newblock {FIRE}: enabling reciprocity for {FDD} {MIMO} systems.
\newblock In \emph{Proceedings of the 27th Annual International Conference on Mobile Computing and Networking}, page 628–641, New York, NY, USA, 2021. Association for Computing Machinery.

\bibitem[Lu et~al.(2024)Lu, Vattheuer, Mirzasoleiman, and Abari]{10.5555/3692070.3693415}
Haofan Lu, Christopher Vattheuer, Baharan Mirzasoleiman, and Omid Abari.
\newblock {NeWRF}: a deep learning framework for wireless radiation field reconstruction and channel prediction.
\newblock In \emph{Proceedings of the 41st International Conference on Machine Learning}. JMLR.org, 2024.

\bibitem[Mildenhall et~al.(2021)Mildenhall, Srinivasan, Tancik, Barron, Ramamoorthi, and Ng]{mildenhall2021nerf}
Ben Mildenhall, Pratul~P Srinivasan, Matthew Tancik, Jonathan~T Barron, Ravi Ramamoorthi, and Ren Ng.
\newblock Nerf: Representing scenes as neural radiance fields for view synthesis.
\newblock \emph{Communications of the ACM}, 65\penalty0 (1):\penalty0 99--106, 2021.

\bibitem[Navabi et~al.(2018)Navabi, Wang, Bursalioglu, and Papadopoulos]{8422221}
Shiva Navabi, Chenwei Wang, Ozgun~Y Bursalioglu, and Haralabos Papadopoulos.
\newblock Predicting wireless channel features using neural networks.
\newblock In \emph{2018 IEEE international conference on communications (ICC)}, pages 1--6. IEEE, 2018.

\bibitem[Park et~al.(2019)Park, Florence, Straub, Newcombe, and Lovegrove]{park2019deepsdf}
Jeong~Joon Park, Peter Florence, Julian Straub, Richard Newcombe, and Steven Lovegrove.
\newblock Deepsdf: Learning continuous signed distance functions for shape representation.
\newblock In \emph{Proceedings of the IEEE/CVF conference on computer vision and pattern recognition}, pages 165--174, 2019.

\bibitem[Radford et~al.(2015)Radford, Metz, and Chintala]{radford2015unsupervised}
Alec Radford, Luke Metz, and Soumith Chintala.
\newblock Unsupervised representation learning with deep convolutional generative adversarial networks.
\newblock \emph{arXiv preprint arXiv:1511.06434}, 2015.

\bibitem[Rappaport et~al.(2013)Rappaport, Gutierrez, Ben-Dor, Murdock, Qiao, and Tamir]{6387266}
Theodore~S. Rappaport, Felix Gutierrez, Eshar Ben-Dor, James~N. Murdock, Yijun Qiao, and Jonathan~I. Tamir.
\newblock Broadband millimeter-wave propagation measurements and models using adaptive-beam antennas for outdoor urban cellular communications.
\newblock \emph{IEEE Transactions on Antennas and Propagation}, 61\penalty0 (4):\penalty0 1850--1859, 2013.

\bibitem[Rappaport et~al.(2015)Rappaport, MacCartney, Samimi, and Sun]{7109864}
Theodore~S. Rappaport, George~R. MacCartney, Mathew~K. Samimi, and Shu Sun.
\newblock Wideband millimeter-wave propagation measurements and channel models for future wireless communication system design.
\newblock \emph{IEEE Transactions on Communications}, 63\penalty0 (9):\penalty0 3029--3056, 2015.

\bibitem[Rappaport et~al.(2017)Rappaport, Xing, MacCartney, Molisch, Mellios, and Zhang]{rappaport2017overview}
Theodore~S. Rappaport, Yunchou Xing, George~R. MacCartney, Andreas~F. Molisch, Evangelos Mellios, and Jianhua Zhang.
\newblock Overview of millimeter wave communications for fifth-generation (5g) wireless networks—with a focus on propagation models.
\newblock \emph{IEEE Transactions on Antennas and Propagation}, 65\penalty0 (12):\penalty0 6213--6230, 2017.

\bibitem[Roelfs and De~Keninck(2023)]{Roelfs_2023}
Martin Roelfs and Steven De~Keninck.
\newblock Graded symmetry groups: Plane and simple.
\newblock \emph{Advances in Applied Clifford Algebras}, 33\penalty0 (3), 2023.

\bibitem[Ruhe et~al.(2023)Ruhe, Gupta, De~Keninck, Welling, and Brandstetter]{10.5555/3618408.3619627}
David Ruhe, Jayesh~K Gupta, Steven De~Keninck, Max Welling, and Johannes Brandstetter.
\newblock Geometric clifford algebra networks.
\newblock In \emph{International Conference on Machine Learning}, pages 29306--29337. PMLR, 2023.

\bibitem[Shen and Wang(2025)]{11206262}
Jingzhou Shen and Xuyu Wang.
\newblock An efficient and explainable kan framework for wireless radiation field prediction.
\newblock In \emph{2025 IEEE 22nd International Conference on Mobile Ad-Hoc and Smart Systems (MASS)}, pages 51--59, 2025.

\bibitem[Shen et~al.(2025)Shen, Zhao, Wu, and Wang]{shen2025nerfaptnewnerfframework}
Jingzhou Shen, Tianya Zhao, Yanzhao Wu, and Xuyu Wang.
\newblock {NeRF-APT}: A new {NeRF} framework for wireless channel prediction, 2025.

\bibitem[Spinner et~al.(2024)Spinner, Bres{\'o}, De~Haan, Plehn, Thaler, and Brehmer]{NEURIPS2024_277628cf}
Jonas Spinner, Victor Bres{\'o}, Pim De~Haan, Tilman Plehn, Jesse Thaler, and Johann Brehmer.
\newblock Lorentz-equivariant geometric algebra transformers for high-energy physics.
\newblock \emph{Advances in neural information processing systems}, 37:\penalty0 22178--22205, 2024.

\bibitem[Wang et~al.(2018)Wang, Bian, Sun, Zhang, and Zhang]{wang2018survey}
Cheng-Xiang Wang, Ji Bian, Jian Sun, Wensheng Zhang, and Minggao Zhang.
\newblock A survey of {5G} channel measurements and models.
\newblock \emph{IEEE Communications Surveys \& Tutorials}, 20\penalty0 (4):\penalty0 3142--3168, 2018.

\bibitem[Wang et~al.(2020)Wang, Wang, Mao, Zhang, Periaswamy, and Patton]{wang2020indoor}
Xiangyu Wang, Xuyu Wang, Shiwen Mao, Jian Zhang, Senthilkumar~CG Periaswamy, and Justin Patton.
\newblock Indoor radio map construction and localization with deep {G}aussian processes.
\newblock \emph{IEEE Internet of Things Journal}, 7\penalty0 (11):\penalty0 11238--11249, 2020.

\bibitem[Wang et~al.(2022)Wang, Wang, Mao, Zhang, Periaswamy, and Patton]{wang2022adversarial}
Xiangyu Wang, Xuyu Wang, Shiwen Mao, Jian Zhang, Senthilkumar~CG Periaswamy, and Justin Patton.
\newblock Adversarial deep learning for indoor localization with channel state information tensors.
\newblock \emph{IEEE internet of things journal}, 9\penalty0 (19):\penalty0 18182--18194, 2022.

\bibitem[Wen et~al.(2025)Wen, Tong, Hu, Lin, and Zhang]{11044513}
Chaozheng Wen, Jingwen Tong, Yingdong Hu, Zehong Lin, and Jun Zhang.
\newblock {WRF-GS}: Wireless radiation field reconstruction with {3D} {G}aussian splatting.
\newblock In \emph{IEEE INFOCOM 2025 - IEEE Conference on Computer Communications}, pages 1--10, 2025.

\bibitem[Wen et~al.(2026)Wen, Tong, Hu, Lin, and Zhang]{wen2024neural}
Chaozheng Wen, Jingwen Tong, Yingdong Hu, Zehong Lin, and Jun Zhang.
\newblock Neural representation for wireless radiation field reconstruction: A 3d gaussian splatting approach.
\newblock \emph{IEEE Transactions on Wireless Communications}, 25:\penalty0 7490--7504, 2026.

\bibitem[Yang et~al.(2025)Yang, Dong, JI, Du, and Srivastava]{yang2025gsrf}
Kang Yang, Gaofeng Dong, Sijie JI, Wan Du, and Mani Srivastava.
\newblock {GSRF}: Complex-valued {3D} {G}aussian splatting for efficient radio-frequency data synthesis.
\newblock In \emph{The Thirty-ninth Annual Conference on Neural Information Processing Systems}, 2025.

\bibitem[Yang et~al.(2019)Yang, Li, Zhang, Qin, Zhu, and Wang]{yang2019generative}
Yang Yang, Yang Li, Wuxiong Zhang, Fei Qin, Pengcheng Zhu, and Cheng-Xiang Wang.
\newblock Generative-adversarial-network-based wireless channel modeling: Challenges and opportunities.
\newblock \emph{IEEE Communications Magazine}, 57\penalty0 (3):\penalty0 22--27, 2019.

\bibitem[Yun and Iskander(2015)]{7152831}
Zhengqing Yun and Magdy~F. Iskander.
\newblock Ray tracing for radio propagation modeling: Principles and applications.
\newblock \emph{IEEE Access}, 3:\penalty0 1089--1100, 2015.

\bibitem[Zhang et~al.(2024)Zhang, Zhan, Xu, Lu, and Xing]{zhang2024fregs}
Jiahui Zhang, Fangneng Zhan, Muyu Xu, Shijian Lu, and Eric Xing.
\newblock Fregs: {3D} {G}aussian splatting with progressive frequency regularization.
\newblock In \emph{Proceedings of the IEEE/CVF Conference on Computer Vision and Pattern Recognition}, pages 21424--21433, 2024.

\bibitem[Zhang et~al.(2025)Zhang, Sun, Berweger, Gentile, and Hu]{zhang2025rf3dgswirelesschannelmodeling}
Lihao Zhang, Haijian Sun, Samuel Berweger, Camillo Gentile, and Rose~Qingyang Hu.
\newblock {RF-3DGS}: Wireless channel modeling with radio radiance field and 3d gaussian splatting, 2025.

\bibitem[Zhao et~al.(2023)Zhao, An, Pan, and Yang]{10.1145/3570361.3592527}
Xiaopeng Zhao, Zhenlin An, Qingrui Pan, and Lei Yang.
\newblock {NeRF2}: Neural radio-frequency radiance fields.
\newblock In \emph{Proceedings of the 29th Annual International Conference on Mobile Computing and Networking}, New York, NY, USA, 2023. Association for Computing Machinery.

\bibitem[Zhao et~al.(2024)Zhao, Wang, An, and Yang]{10.1145/3678572}
Xiaopeng Zhao, Shen Wang, Zhenlin An, and Lei Yang.
\newblock Crowdsourced geospatial intelligence: Constructing {3D} urban maps with satellitic radiance fields.
\newblock \emph{Proc. ACM Interact. Mob. Wearable Ubiquitous Technol.}, 8\penalty0 (3), 2024.

\end{thebibliography}
}

\end{document}